\documentclass[aps,prb,groupedaddress,twocolumn]{revtex4}

\usepackage{graphicx}
\usepackage{amssymb,amsmath}
\usepackage{subfigure}
\usepackage{epstopdf}

\begin{document}


\title{Dynamics of a Polariton Condensate in an Organic Semiconducting Microcavity.}


\author{Eric R. Bittner}
\email[]{Bittner@UH.edu}
\homepage[]{http://k2.chem.UH.edu}
\affiliation{Department of Chemistry, University of Houston, Houston, TX 77204}

\author{Svitlana Zaster}
\affiliation{Department of Chemistry, University of Houston, Houston, TX 77204}

\author{Carlos Silva}
\affiliation{ Department of Physics and 
Regroupement Qu{\'e}b{\'e}cois sur les Mat{\'e}riaux de Pointe, Universit{\'e} de Montr{\'e}al,\\
C.P. 6128, Succursale centre-ville, \\
Montr{\'e}al (Qu{\'e}bec) H3C 3J7, Canada.}


\date{\today}

\begin{abstract}

 Recent experiments on thin-film microcavities give evidence of 
Bose condensation of exciton-polariton states. Inspired by these observations, we 
consider the possibility that such exotic ``half-light/half matter" states could be observed
in thin-film organic semiconductors where the oscillator strength is generally stronger than in 
inorganic systems.   Here we present a theoretical model  and simulations of 
 macroscopic exciton-polartiton condensates in  anthracene thin films sandwiched within a micro-meter scale resonant cavity and 
 establish criteria for the conditions under which BEC could be achieved in these systems.
 We consider the effect of lattice disorder on the threshold intensities necessary to create polartion superfluid states
 and conclude that even allowing for up to 5\% angular disorder of the molecules within the crystal lattice, 
 the superfluid transition remains sharp. 
\end{abstract}


\maketitle

\section{Introduction}
Over the past few years there has been considerable interest in the dynamics of cavity polaritons in low-dimensional 
inorganic microcavities.
\cite{Amo:2009fk,Bajoni:2007pb,Balili:2007ai,Kasprzak:2006mz,Lagoudakis:2008ij,Malpuech:2007ec,Kasprzak:2008mi,PhysRevLett.102.056402}
In the strong coupling regime, 
where the photon-exciton interaction is larger than the exciton and photon damping rates, 
photons within the cavity and excitons within the material become coherently coupled to form bosonic quasi-particles 
termed ``polaritons''.   \cite{agranovich:075302,D.-D.-C.-Bradley:1998xw,littlewood:15} 
Macroscopic Bose-Einstein condensate (BEC) states have recently been reported in GaAs and in CdTd/CdMgTe microcavities\cite{Kasprzak:2006mz,littlewood:15,tsintzos:071109,wertz:051108} and recent reports   give evidence of polariton condensation in one-dimensional ZnO microwires.
\cite{Sun:2010uq}

Polaritons obey a dispersion  relation that splits into two branches with an avoided crossing near the intersection of the cavity dispersion and the exciton resonance frequency.  
In a typical inorganic quantum-well semiconductor microcavity, the Rabi splitting between  Wannier-Mott excitons and the cavity modes is typically on the order of 10 meV. However, for organic systems (with Frenkel excitons) exciton oscillator strengths are considerably higher than their inorganic counterparts and consequently the Rabi splittings can be on the order of 100 meV.   \cite{Kena-Cohen:2008tw,PhysRevB.71.235203}  
While polariton lasing has been reported in organic microcavities,\cite{Kena-CohenS.:2010fk} thus far, polariton BEC has not been in these systems. 
This paper is also motivated by the recent reports of BEC of a two-dimensional photon gas in a dye-filled optical microcavity.\cite{Klaers:2010fk}  In these experiments, the 
strong coupling requirement is not met and the quasi-particles within the cavity are indeed best described as photons rather than polaritons.  

In this paper, we consider the preparation and dynamics of BEC states in an organic microcavity consisting of an acene thin-film sandwiched within a microcavity of thickness $L$. 
To date, the evidence for BEC has not been reported in organic semiconductor microcavity systems.  Whereas in inorganic systems, the 
electronic states are best described as Wannier excitons with weakly bound electron-hole pairs, 
in organic semiconductors the excitons are Frenkel excitons that are largely localized on  a single molecular unit.  Consequently, their
electronic transition moments are generally larger compared to inorganic systems resulting stronger coupling to the photon field 
and consequently larger Rabi splittings.  

\begin{figure}[hb]
\includegraphics[width=\columnwidth]{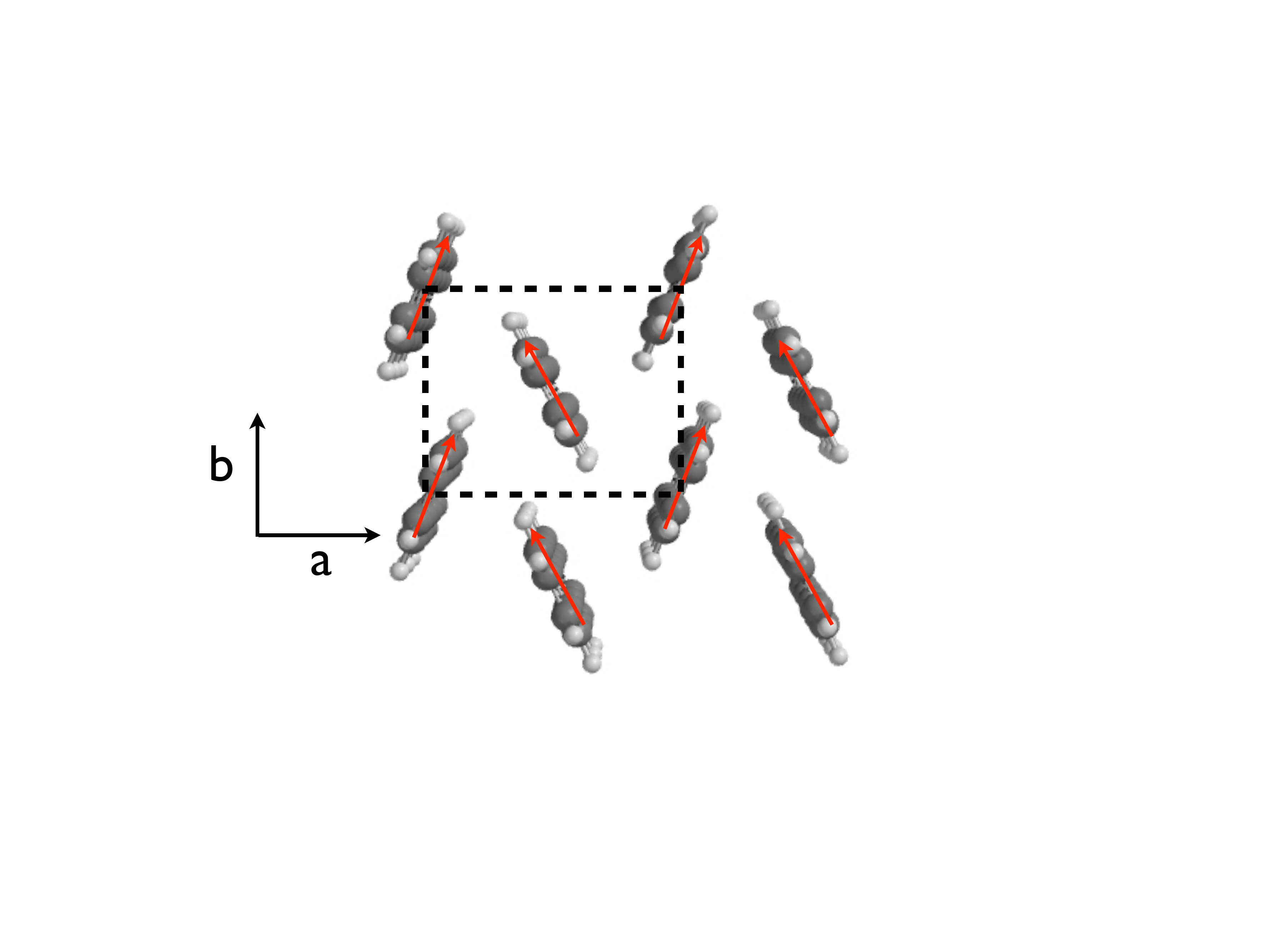}
\caption{Anthracene unit-cell in the $\{{\rm a},{\rm b}\}$ crystalographic plane.  Arrows indicate the direction of the 
$S_o\to S_1 $ electronic transition dipole moment.}\label{unit-cell}
\end{figure}

While BEC in atomic systems involves weakly interacting particles, these systems are not uniform since the condensate is influenced by the form of the trapping potential.
While liquid $^{4}$He, for example, is also a BEC condensate, perturbative treatments can not be used on this system due to strong inter-particle interactions and correlations.    
Consequently, polaritons offer a unique glimpse into BEC physics since the shape of the condensate state itself is {\em unambiguous} for a uniform system. 
The specific signature of a BEC state in the form of a spontaneous coherence offers a unique opportunity to 
 study localization effects in two-dimensional systems with varying degrees of spatial disorder. 

 In a cavity system, the disorder can be introduced either by speckles within the optical field,  by the orientational disorder of the molecular sites, 
 or by site-energy disorder within a perfectly periodic crystal structure.  
 The role of disorder is an interesting issue since it would allow one to probe indirectly at least 
 the transition between the superfluid and Mott insulating regime.  There has been some speculation that a Bose-glass phase is intermediate 
 between the superfluid  and Mott insulating phases implying that disorder transforms the Mott insulting phase to a Bose-glass. However, 
 other theories suggest that disorder can change the Mott insulating phase directly to a superfluid. Recent experimental evidence 
 involving trapped atoms indicate the transition is disorder-induced  and reversible for superfluid and coexisting superfluid-Mott insulator phases. 
Furthermore, spatial disorder plays a key role in the fractional quantum Hall effect  and vortex pinning in type-II superconductors.  

Our analysis begins with a molecular-level description of a single layer of polyacene molecules in which the local electronic 
excitations are described within the Frenkel exciton model with dipole-dipole couplings determined by quantum chemistry. 
The local excitons interact with a single mode of the photon field within the dipole approximation and we determine
the polariton dispersion curves by diagonalizing the full exciton+photon Hamiltionian for a given configuration of the lattice molecules.
 This treatment allows us to take an arbitrary configuration of the molecules within the lattice and generate 
polariton dispersions.   In Sec. III we adopt a field-theoretical approach developed by Carusotto and Ciuti which treats the cavity as a non-equilibrium steady-state system.  \cite{Carusotto:2004sp,Carusotto:2006la,PhysRevA.80.043603,PhysRevB.72.125335,PhysRevB.76.115324,PhysRevLett.87.064801}
This allows us to identify steady-state polaritons as parameterized by the intensity of the driving field.  We then use the Bogoliubov prescription to separate fluctuations from  the stationary solution and use this to determine the excitation spectra for the lower polariton branch.   We then consider the response of the system to inhomogeneous perturbations within the cavity and show that above a threshold intensity for the driving field, the scattering signal collapses to a single point indicating the onset of superfluidity. 

The cut-off temperature for the amplification is ultimately determined by the binding energy of the exciton.
 Because this is large in organic semiconductors (hundreds of meV), parametric 
  amplification should be readily observable at significantly higher temperatures 
  than in inorganic quantum-well microcavities, even at room temperature.

\section{Theoretical model}

\begin{figure}[t]
\subfigure[J-aggregate]{\includegraphics[width=0.48\columnwidth]{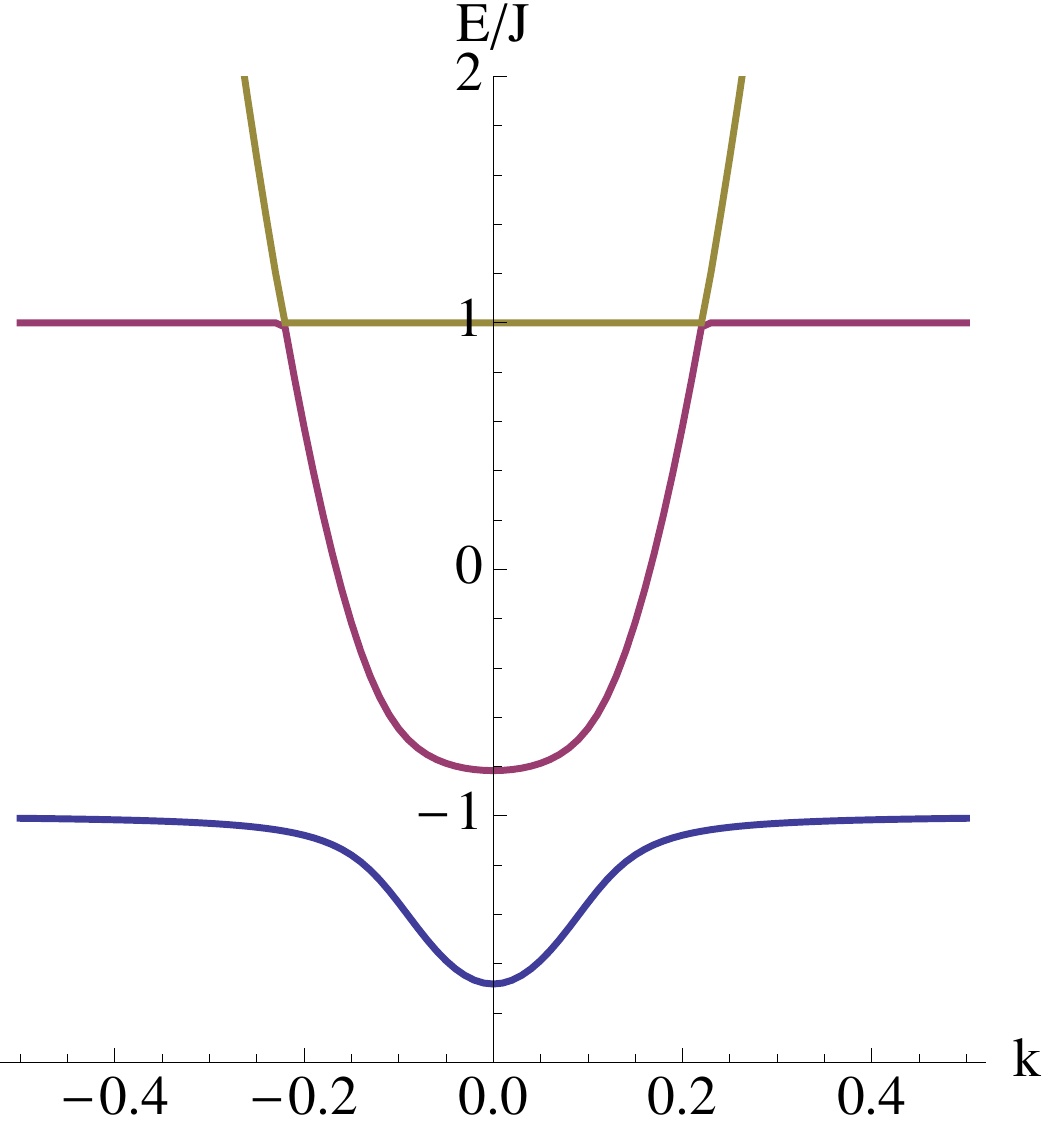}}
\subfigure[H-aggregate]{\includegraphics[width=0.48\columnwidth]{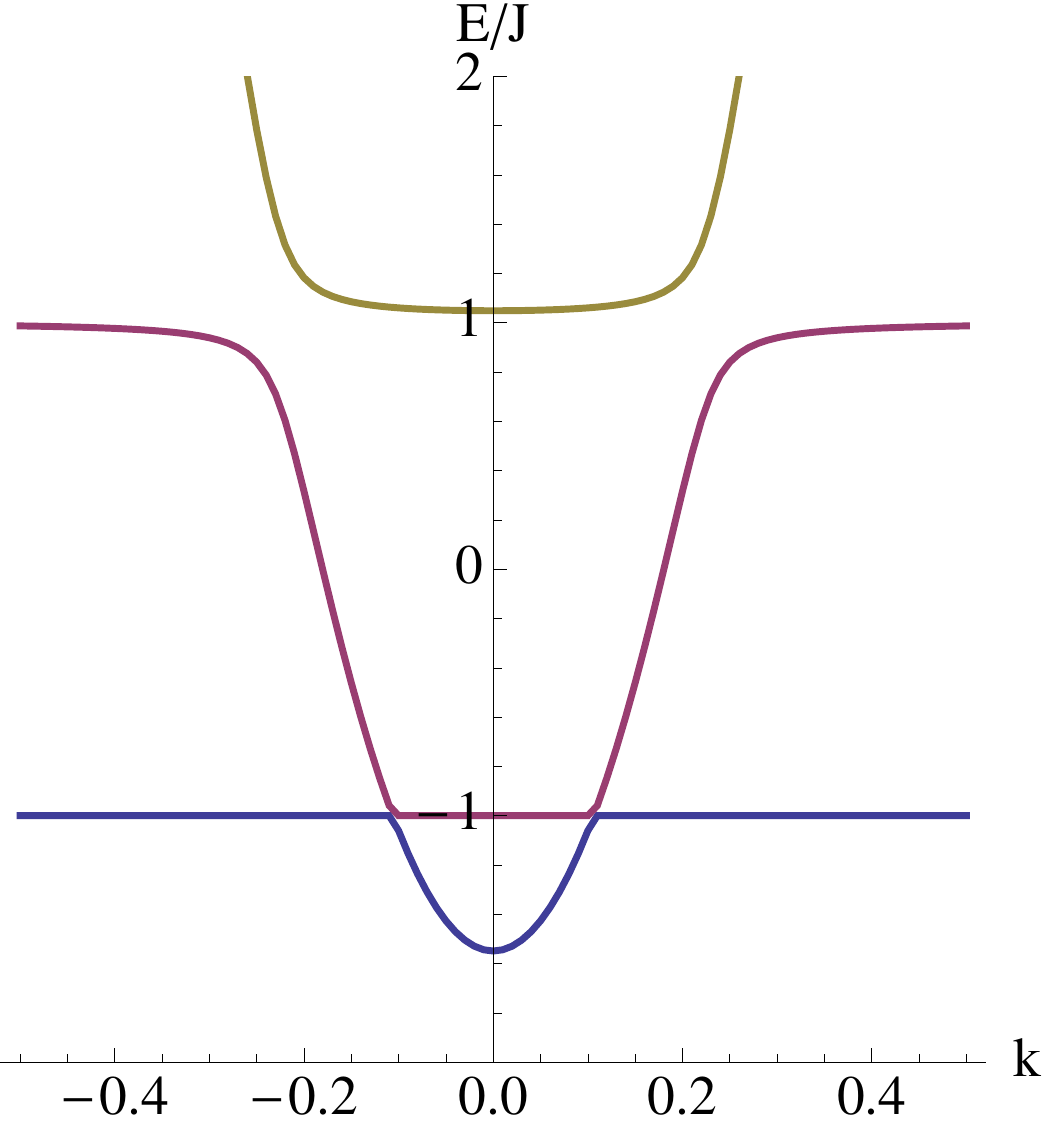}}
\caption{Polariton/exciton dispersion curves for (a) co-parallel ($\leftarrow \cdots\leftarrow)$ and (b) co-facial ($\uparrow \cdots\uparrow)$ aligned dipoles.}\label{aggs}
\end{figure}

Polaritons are composite quasi-particles that form due to the strong coupling between excitons and the radiation field.  
Denote $\hat a^\dagger$ as an operator that creates an excitation with energy $\epsilon_{x}$ 
and $\hat a$ as an operator which removes an excitation 
with $\{\hat a,\hat a^{\dagger}\} = 1$  and  with similar operators for the radiation field, $[\hat\psi_{k},\hat\psi_{k'}^{\dagger}] = \delta_{kk'}$
which create and remove photons with wave-vector $k$. 
The Hamiltonian describing the coupled exciton/photon system is given by 
\begin{eqnarray}
H = \epsilon_{x}\hat a^{\dagger} \hat a + \sum_{k} \hbar\omega_{k}\hat\psi^{\dagger}_{k}\hat\psi_{k} +\frac{ \Omega_{R}}{2} (\hat a^{\dagger}\hat \psi_{k}+  \hat \psi_{k}^{\dagger}\hat a)
\end{eqnarray}
where $\Omega_{R}$ is the Rabi frequency which depends  upon the number of photons in a given mode $k$, the oscillator strength of the transition, 
and the orientation of the excitation's transition moment relative to the polarization of the photon. 
Close to the resonance condition $\hbar\omega_{k} = \epsilon_{x}$, the energy eigenstates of system are best described 
in a mixed or entangled representation of the system given by 
\begin{eqnarray}
\left(\begin{array}{c} \hat a^{\dagger} \\\hat\psi_{k}^{\dagger}  \end{array}\right)
= 
\left(
\begin{array}{cc}
\cos\theta_{k}  & \sin\theta_{k} \\
-\sin\theta_{k}  & \cos\theta_{k} 
\end{array}
\right)
\left(\begin{array}{c}\hat L_{k}^{\dagger} \\ \hat U_{k}^{\dagger} \end{array} \right)
\end{eqnarray}
where the mixing angle is given by $\tan 2\theta_{k} = \Omega_{R}/(\epsilon_{x} - \hbar\omega_{k})$ where the $U_{k}^{\dagger}$ and $L_{k}^\dagger$ 
operators are  quasi-particle operators which create polaritons in the upper (U) or lower (L) branches with energy
$$
E_{k}^{\pm} = \frac{\epsilon_{x} + \hbar\omega_{k}}{2}\mp \sqrt{\left(\frac{\epsilon_{x} -\hbar\omega_{k}}{2} \right)^{2}+ \hbar^{2}\Omega_{R}^{2}}
$$
where $+$ denotes the upper polariton (UP) and $-$ denotes the lower polariton (LP) branches respectively.


 \begin{figure}[t]
\subfigure[]{\includegraphics[width=0.49\columnwidth]{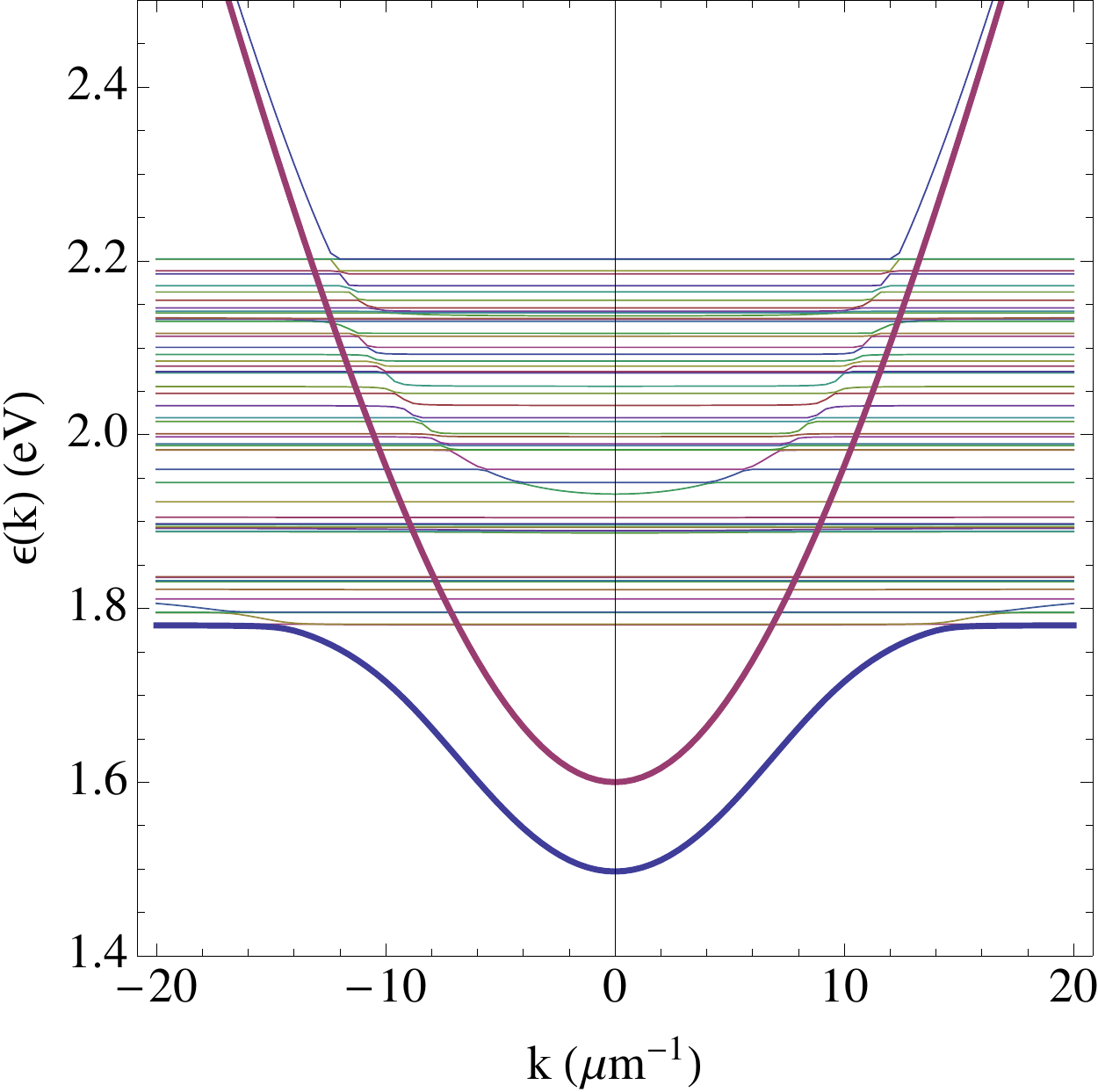}}
\subfigure[]{\includegraphics[width = 0.49 \columnwidth]{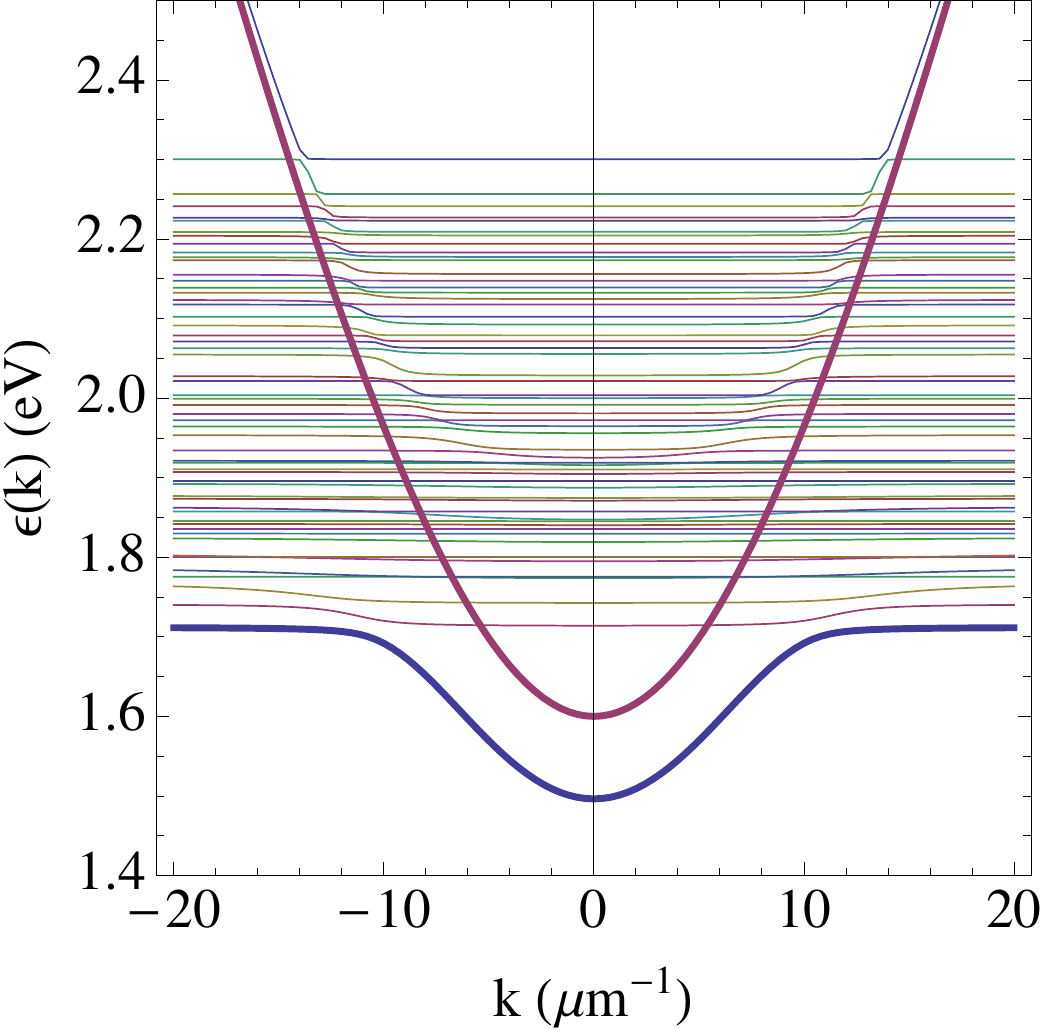}}
\subfigure[]{\includegraphics[width=0.49\columnwidth]{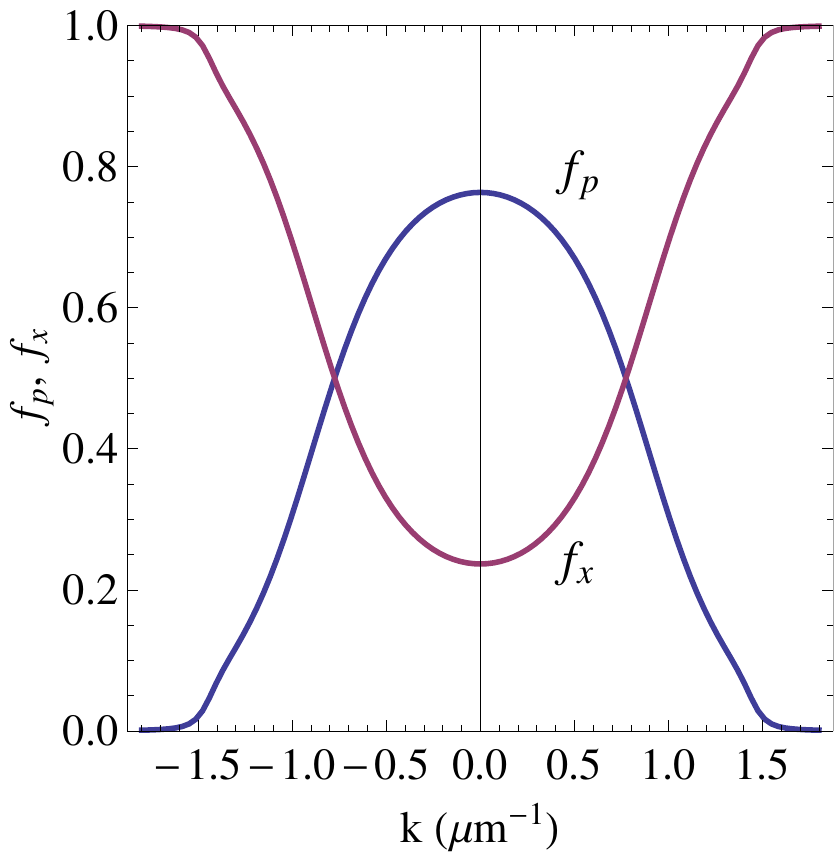}}
\subfigure[]{\includegraphics[width=0.49\columnwidth]{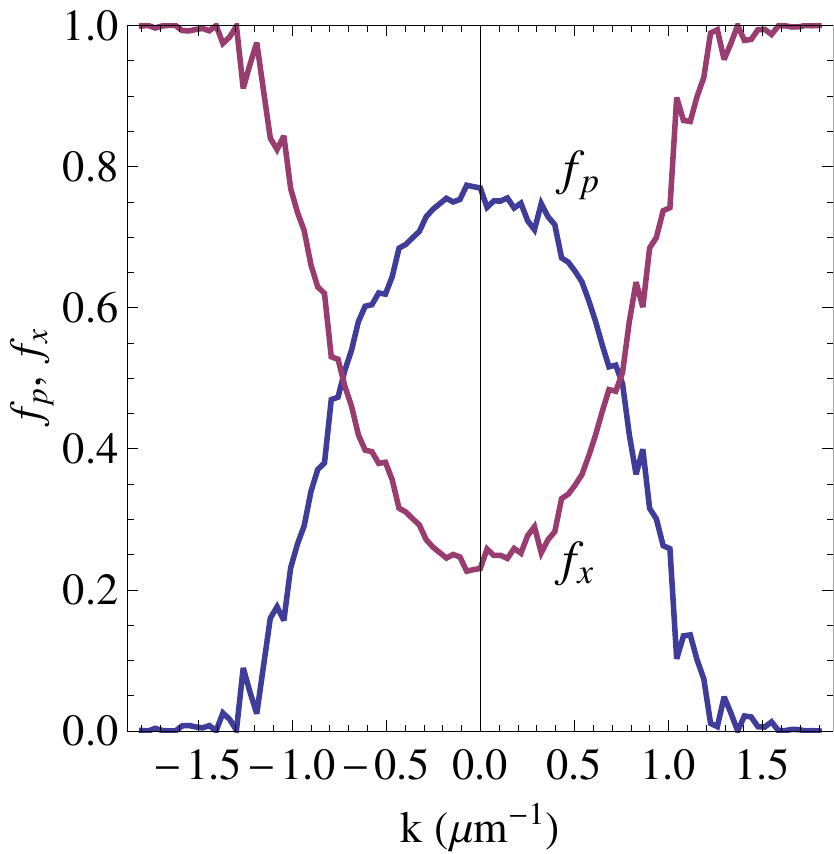}}
\caption{ Polariton energy levels vs. photon wave-vector for a finite anthracene slab. 
The thick blue curve corresponds to the {\em lower polariton} branch (LP) while the thick red curve is the cavity dispersion.
 At  two upper polariton branches can as well as evidenced by the
avoided crossings within the exciton bands.  
(a) Exciton-polariton dispersion for perfectly ordered anthracene lattice.
(b)Exciton-polariton dispersion when $\delta\bar\theta =15^{\circ}$ 
of orientational disorder is included in the slab model. 
(c,d)Exciton $f_{x}$ and photon $f_{p}$ fraction for dispersions in (a,b)} \label{dispersion1}
\end{figure}

Within a microcavity, the photons obey standing-wave boundary conditions such that the transverse modes are quantized
 giving rise to the dispersion 
\begin{eqnarray} \epsilon_k=(\hbar c/\eta)\sqrt{k^{2} + (2\pi n/L)^{2}}\end{eqnarray}
  where $\eta$ is the refractive index
and $n = 1,2,\cdots$ is the index of the transverse mode.  About $k = 0$, the cavity dispersion is quadratic 
\begin{eqnarray} \epsilon_{k} \approx \Delta + \frac{\hbar^{2}k^{2}}{2m^{*}} + \cdots\end{eqnarray}
where
$m^{*} = 2\pi\eta/cL$
 is the effective mass of the cavity photon 
and $\Delta =  2\pi \hbar c/L \eta$ is the cut-off energy for the cavity.
 From this point onward, operators $\hat\psi_{\bf k}^\dagger$ and $\hat \psi_{\bf k}$ create and remove photons
 within the microcavity and index ${\bf k}$ refers to the photon wavevector in the plane of the slab.

We can extend this simple model to describe the interaction between a photon field and slab of 
molecular sites within the cavity.  First, we assume that the excitons are local to the molecular sites and  can be described within the
 Frenkel exciton model using the Hamiltonian\cite{agranovich:075302}
\begin{eqnarray}
H_{ex} &=& \sum_{i}\epsilon_{i}\hat a_{i}^{\dagger}\hat a_{i} + \sum_{i\ne j}J_{ij}(\hat a_{i}^{\dagger}\hat a_{j} + \hat a_{j}^{\dagger}\hat a_{i}) 
\label{Hsite}
\end{eqnarray} 
 where the operators $\hat a^{\dagger}_{i}$ and $\hat a_{i}$ create and remove 
 excitons at site $i$ located at ${\bf r}_{i}$.   
 In the absence of internal relaxation processes, 
 the exciton dynamics within the film are determined by the exchange integral $J_{ij}$.
Within the point-dipole approximation, this can be written as 
\begin{eqnarray}
J_{ij} =\frac{1}{\eta^2} \left(
\frac{3 (\mu_{i}\cdot {\mathbf r_{ij}})(\vec\mu_{j}\cdot {\mathbf r_{ij}})}{r_{ij}^{5}}-\frac{\vec\mu_{i}\cdot \mu_{j}}{r_{ij}^{3}}\right)
\end{eqnarray}
where $\vec\mu_{i}$ are 
transition moments for the $S_{o}\to S_{1}$ electronic transition on a 
given site  and ${\mathbf r_{ij}} = {\mathbf r}_{i}- {\mathbf r}_{j}$ is the vector
connecting the centers of molecules $i$ and $j$.  
Fig.~\ref{unit-cell} shows the  ``herringbone'' arrangement of the anthracene
 molecules within a single crystallograpic plane.

The coupling between the cavity modes and the excitons is given by 
\begin{eqnarray}
\hat\Omega/2 = \gamma \sum_{i{\bf k}}\vec\epsilon\cdot\vec\mu_i\sqrt{\frac{\epsilon_{i}}{N\epsilon_{{\bf k}}}}(e^{i{\bf {\bf k}}\cdot{\bf x}_{i}}\hat a_{i}^{\dagger}\hat\psi_{{\bf k}} +e^{-i{\bf {\bf k}}\cdot{\bf x}_{i}}\hat\psi_{{\bf k}}^{\dagger}\hat a_{i})\label{rabiop}
\end{eqnarray}
where $N$ is the number of sites within the 2D slab and $\vec\epsilon$ is the polarization vector of the photon. 
  The model used here has been used in  a variety of contexts, notably by Agranovich and Gartstein
 in their recent study of low-energy exciton-polaritons in one-dimensional systems.  \cite{agranovich:075302}
 The site-model is particularly attractive since it allows us to parameterize it using both spectroscopic and 
 quantum-chemical information about a specific molecular system. 
 
  One should, in principle, add to our model the fact that the excitons are composite quasi-particles
 consisting of bound electron/hole pairs.   For organic systems, the exciton binding energy is in the range of $0.4$ to $0.5$eV
 and the electron and hole largely reside on the same molecular site, which justifies the use of the Frenkel exciton model for these systems. 
\cite{Rashba:1963,Spano:2010xc,spano:5376}
An important feature which we do neglect, however, is the coupling between the local site exciton and the internal vibrational motions 
of the molecule itself. In organic systems, such vibronic couplings play important roles in the spectroscopy and 
dynamics of conjugated systems.\cite{Spano:2010xc,bittner:57,karabunarliev:3988,karabunarliev:4291,karabunarliev:5863}

Since we are interested in polaritons in molecular crystals,  we need to consider how the crystallographic arrangement 
and possible molecular disorder will affect the formation of polariton states within such a system.  With this in mind we 
consider an ordered 2D array of acene molecules within a microcavity.
At 290 K, polyacene systems typically form monoclinic crystals with
 two molecules per unit cell in the plane containing the molecular $y$ axis oriented
alternately $\pm 20^{\circ}$ from the crystallographic $b$ direction.  When viewed from above the plane, we see the ``herringbone'' 
 packing geometry characteristic of polyacene molecular crystals. The electronic transition moment for the $S_{o}\to S_{1}$ transition lies along the molecular $y$ axes
and hence lies in a plane parallel to the reflectors.

Consider the case of a molecular dimer (such as two polyacenes)
interacting with a photon field where we let  $\hat a_{1}$ and $\hat a_{1}^{\dagger}$ denote exciton operators
for one monomer and $\hat a_{2}$ and $\hat a_{2}^{\dagger}$ be exciton operators for the other with the exchange interaction denoted by $J$. 
In the ``local'' $\{\hat\psi_k,\hat a_{1},\hat a_{2}\}$ basis,  we can write
\begin{eqnarray}
H &=& 
\hbar\omega_{k} \hat\psi_{k}^{\dagger}\hat\psi_{k}   
+ \epsilon_{1}\hat a_{1}^{\dagger} \hat a_{1}
+ \epsilon_{2}\hat a_{2}^{\dagger} \hat a_{2} \nonumber \\
&+& J (\hat a_{1}^{\dagger} \hat a_{2} + \hat a_{2}^{\dagger} \hat a_{1}) \\ 
&+&\Omega_{R}/2 (\hat\psi_{k}^{\dagger}(\hat a_{1} + \hat a_{2}) + h.c.) 
\end{eqnarray}
Taking $\epsilon_{1} = \epsilon_{2}$, $H$ can be brought into block-diagonal form 
by creating Davydov exciton states via
\begin{eqnarray}
\hat D_{\pm}  = (\hat a_{1} \pm \hat a_{2})/\sqrt{2}
\end{eqnarray}
which results in
\begin{eqnarray}
\tilde H  &=& 
\hbar\omega_{k} \hat\psi_{k}^{\dagger}\hat\psi_{k}   
+ (\epsilon_{1}+J) \hat D_{+}^{\dagger} \hat D_{+} \nonumber \\
&+& (\epsilon_{1}-J) \hat D_{-}^{\dagger} \hat D_{-}\nonumber \\
&+&\Omega_{R} (\hat\psi_{k}^{\dagger}\hat D_{+} + \hat D_{+}^{\dagger}\hat\psi_{k}) /2
\end{eqnarray}
Taking the transition moments parallel in the same plane where $\theta$ is  the angle between the transition moment and the 
centers of the two molecules, i.e. $\nearrow\cdots\nearrow$,
 the exchange coupling is given by 
\begin{eqnarray}
J =-\frac{\mu^2 }{R^3 \eta}\left(2 \cos ^2(\theta )-\sin ^2(\theta )\right) 
\end{eqnarray}
For $0< \theta < \cos^{-1}(1/\sqrt{3})$, $J< 0$ and the dimers form a J-aggregate while for angles $\cos^{-1}(1/\sqrt{3})< \theta < \pi/2$, $J>0$ and the 
dimers form an H aggregate.  At $\theta = \cos^{-1}(1/\sqrt{3}) =54.7^{\circ}$, the exchange coupling vanishes and no exciton splitting occurs.

We now note that only the {\em symmetric} Davydov state is coupled to the photon field and we can use a unitary transformation
to define the upper and lower polariton branches 
\begin{eqnarray}
\left(\begin{array}{c} \hat D_{+} \\\hat\psi_{k}  \end{array}\right)
= 
\left(
\begin{array}{cc}
\cos\phi_{k}  & \sin\phi_{k} \\
-\sin\phi_{k}  & \cos\phi_{k} 
\end{array}
\right)
\left(\begin{array}{c}\hat L_{k} \\ \hat U_{k} \end{array} \right)
\end{eqnarray}
where the mixing angle is given by $\tan 2\phi_{k} = \Omega_{R}/(\epsilon_{1} - \hbar\omega_{k})$ where the $U_{k}^{\dagger}$ and $L_{k}^\dagger$ 
operators are  quasi-particle operators which create polaritons in the upper (U) or lower (L) branches 
with dispersion
\begin{eqnarray}
E_{k}^{\pm} = \frac{(\epsilon_{1}+J) + \hbar\omega_{k}}{2}\mp \sqrt{\left(\frac{(\epsilon_{1}+J) -\hbar\omega_{k}}{2}\right)^{2} + \hbar^{2}\Omega_{R}^{2}}
\label{davidov-polariton}
\end{eqnarray}
Consequently, for a molecular dimer, we have three states: the upper and lower polarition states and a dark exciton state. 
The relative ordering of the exciton energy levels depends upon the sign of $J$, which in turn depends upon angle between 
the co-facial stacking planes.   For J-aggregates, $J< 0$ the bright state is lower in energy than the 
dark state where as for H-aggregates $J>0$ and the dark exciton state is lower in energy.
In Fig.~\ref{aggs} we compare the dispersion curves for a molecular dimer taken as a J-aggregate with $\theta = 0$
and an H-aggregate ($\theta = \pi/2$).  Here we use parameters: $\hbar\omega_{k}  = k^{2}/2m^{*} + \Delta$, $m^{*} = 0.01$,  $\Omega_{R} = 0.5 |J|$, and $\Delta = 0$ 
for purposes of this discussion.
For the J aggregate, the lower Davydov state carries a net coupling to the photon mode and the lower polariton branch is formed solely 
by coupling to this state. The upper Davydov state does not contribute the formation of polaritons.   The situation is exactly reversed for the H-aggregates
in that only the upper Davydov state is coupled to the photon field. 

This basic idea can be expanded to describe the full 2D slab in two ways.  First, we can write the transition moment for the unit cell by 
taking the vector sum and differences
$
\vec \mu_{\pm} = (\vec\mu_{1}\pm \vec\mu_{2})/\sqrt{2}
$
of the molecular transition moments within the cell. 
This results in  transition moments $\mu_{\pm}$ which are orthogonal and directed along the $a$ and $b$ crystallographic  axes.  Thus, 
taking the dipole-dipole coupling to be short ranged so that only neighboring cells are coupled, the exciton dispersion can be written 
as
$$
E_{x}(k) = \epsilon_{o} + 2J_{a} \cos(k_{x} x) + 2 J_{b}\cos(k_{y}y)
$$
where $J_{a}$ and $J_{b}$ are computed using $J_{a} ={\mu_{-}^2 }/{a^3 \eta}$ and $J_{b} = {\mu_{+}^2 }/{b^3 \eta}$ where $a$ and $b$ 
are the lattice constants.  For anthracene: $a = 8.5262$\AA\ and   $b = 6.038$\AA\ are the lattice constants for the planar array shown in Fig. 1. 
Since the wavevector $k$ refers to the wavevector in the lattice plane, this dispersion is flat compared to the 
dispersion for the cavity photons and one can immediately extend Eq.~\ref{davidov-polariton} for the 2D lattice.  

A more realistic approach is to include orientational and energetic disorder directly into the model and 
obtain polariton states by numerical diagonalisation.  
In the calculations presented next, each molecular site is described by a transition moment $\mu_{i} = 1.1901 a.u.$ corresponding to an oscillator 
strength of $f_{osc} = 0.1555$ as determined by {\em ab initio} (CIS/6-31G*) calculations on an isolated anthracene molecule using Gaussian '03.\cite{g03}
Rather than using the {\em ab initio} vertical excitation energy for anthracene, we set the site excitation energy as $\epsilon_{i} = 2.0{\rm eV}$.
In the end, this value simply establishes a reference point and all excitations and energies can be re-scaled relative to the 
vertical exciton energy for purposed of comparison with experiments.
Each dipole is located at the crystallographic site and oriented alternately $\pm 20^{\circ}$ from the crystallographic $b$ direction.  
This coupling gives rise to exciton couplings of $-96.2$meV between the two acenes within the unit cell. We also assume a 
static dielectric constant of $\eta = 3$, a value typical for organic solids.   Finally, we model the cavity by taking a photon energy cut-off of 
 $\Delta = 1.6$eV which is detuned from the exciton resonance energy. 
Using this approach, we can easily include the effects of energetic and orientational disorder on the exciton  dispersion.  
Thus, armed with the molecular crystal geometry, the excitation energies, and the electronic transition moments, one can 
determine the polariton energies varying with photon wave-vector for a given microcavity, by diagonalizing the above Hamiltonian for a given photon 
wave-vector.

In Fig.~\ref{dispersion1}a we show the single-particle eigenvalues of the coupled exciton/photon 
system for photon $k$-vectors between $\pm 25 \mu{\rm m}$ as determined by diagonalizing 
$$\hat H = \hat H_{ex} + \hat H_{ph} + \hat H_{ex/ph}$$ at specific values of $k$.  
Close to $k = 0$, the lower polariton branch is nearly parabolic and one see clear evidence of 
anti-crossings.  The lower thick-line highlights  the ``lower polariton'' (LP) branch in this system. 
 Looking closely, one can identify multiple upper polariton bands that pass through the density of states. 
 Also it is important to note that not every  exciton state couples to the photon field. These dark states 
 result from the fact that their net transition dipole is either very small or nearly perpendicular 
 to the polarization vector of the driving laser field.  
 
Within an organic molecular crystal, orientational disorder of the individual molecules inhibits the 
formation of delocalized excitonic states within the lattice.  Consequently, this may limit the 
ability of the system to form collective coherent excitations such as polaritons.  We can easily examine the effect of
orientational disorder by randomly orienting the dipoles about some mean angle $\bar\theta  = 20^{\circ}$ with 
variance $\sigma_{\theta}^{2}$.  In Fig.\ref{dispersion1}b  we show the polariton curves 
for a slab with  $\sigma_{\theta}= 15^{\circ}$ of orientational disorder. 
The primary effect of the orientational disorder appears to be that polaritons formed by coupling to the lowest 
Davydov states splits into multiple polariton states as evidenced by the increased number of 
anti-crossings that can be spotted within the density of states.  Moreover, the first anti-crossing
occurs at slightly lower values of $k$.   
In Fig.~\ref{dispersion1}(c,d) we show the relative mixing between the cavity photon and the  anthracene slab for both the ordered  slab (c) and the
orientationally disordered slab (d).  In both cases the mixing is quite strong even as one moves from the anti-crossing region
in spite of the fact that the cavity cut-off is $\approx 0.2$eV off resonance from the exciton band. 

\section{Field theory treatment}

While a fully ``molecular'' treatment is possible in principle using a site-wise representation, the $\sim1 \mu m$ length-scales suggested by the 
dispersion curves in Fig.~\ref{dispersion1}  would require two dimensional slabs with upwards of $10^{6}$ anthracene sites.  
On the other hand, the bare exciton dispersions are essentially flat over the range of wave vectors that are important for polariton 
formation within the cavity.   Bearing this in mind, let us define field operators $\hat\psi_{x}$ and $\hat\psi_{c}$ for the exciton and cavity modes
and write the Hamiltonian density within the Hartree-Bose approximation:
\begin{widetext}
\begin{eqnarray}
{\cal H} &=& \int d{\bf x }
\left\{\left(
\begin{array}{c}
\hat\psi_{x}^{\dagger}({\bf x}) \\ \hat\psi_{c}^{\dagger}(\bf x)
\end{array}
\right)^{T}
\cdot
\left(
\begin{array}{cc}
\hat\omega_{x}- i \gamma_{x}/2  + g\hat\psi_{x}^{\dagger}({\bf x})\hat\psi_{x}({\bf x})/2 & \hat\Omega/2\\
\hat\Omega/2  &\hat \omega_{c} - i \gamma_{c}/2 
\end{array}
\right)
\cdot
 \left(
\begin{array}{c}
\hat\psi_{x}({\bf x})\\ \hat\psi_{c}(\bf x)
\end{array}
\right)\right\} \nonumber \\
&+& 
\int d{\bf x}\left( E({\bf x},t)\psi^{\dagger}_{c}({\bf x}) + E^{*}({\bf x},t)\psi_{c}({\bf x}) \right)
\end{eqnarray}
where $\hat\omega_{c}$ and $\hat\omega_{x}$ generate the dispersion curves for the cavity and exciton modes 
and $\gamma_{c}$ and $\gamma_{x}$ describe the cavity and exciton decay rates and $\hat\Omega$ is the Rabi operator 
which couples the cavity field (at a given photon ${\bf k}$-vector) to the exciton field as given by Eq.~\ref{rabiop}.

Steady state solutions for the fields can determined by first writing $E(x,t) = E_{p}(x)\exp(-i\omega_{p}t)$ and $\hat \psi_{\mu}(t) = \phi_{\mu,s}\exp(-i\omega_{p}t)$
and writing 
\begin{eqnarray}
\left(
\begin{array}{cc}
\hat\omega_{x}- i \gamma_{x}/2  + g|\phi_{x,v}|^{2}/2 - \omega_{p}& \hat\Omega/2\\
\hat\Omega/2  &\hat \omega_{c} - i \gamma_{c}/2  - \omega_{p}
\end{array}
\right)
\cdot
 \left(
\begin{array}{c}
\phi_{x,s}({\bf x})\\ \phi_{c,s}(\bf x)
\end{array}
\right)
+
\left(
\begin{array}{c}
0  \\  E_{p}({\bf x}) 
\end{array}
\right)
 = 0.
\end{eqnarray}
This leads to a set of equations for the steady-state,
\begin{eqnarray}
 \left(
\begin{array}{c}
\phi_{x,s}\\ \phi_{c,s}
\end{array}
\right)
=-
\left(
\begin{array}{cc}
\hat\omega_{x}- i \gamma_{x}/2  + g|\phi_{x,v}|^{2}/2 - \omega_{p}& \hat\Omega/2\\
\hat\Omega/2  &\hat \omega_{c} - i \gamma_{c}/2  - \omega_{p}
\end{array}
\right)^{-1}
\cdot
\left(
\begin{array}{c}
0  \\  E_{p}({\bf x}) 
\end{array}
\right).
\end{eqnarray}
These can be solved iteratively for a given driving field, $E_{p}(\bf x)$.  Furthermore, if one assume that the
system is homogeneous and the laser field consists of a single mode at ${\bf k}_{p}$, the steady state equations 
simplify to 
\begin{eqnarray}
 \left(
\begin{array}{c}
\phi_{x,s}\\ \phi_{c,s}
\end{array}
\right)
=-
\left(
\begin{array}{cc}
\omega_{x}({\bf k}_{p})- i \gamma_{x}/2  + g|\phi_{x,v}|^{2}/2 - \omega_{p}&\Omega({\bf k}_{p})/2  \\
\Omega({\bf k}_{p})/2  & \omega_{c}({\bf k}_{p})-  i \gamma_{c}/2  - \omega_{p}
\end{array}
\right)^{-1}
\cdot
\left(
\begin{array}{c}
0  \\  E_{p}
\end{array}
\right).
\end{eqnarray}
where the $\omega_{x}$ and $\omega_{c}$ are the exciton and cavity dispersions evaluated at ${\bf k}_{p}$ of the driving field. 
\end{widetext}

\subsection{Lower Polariton Branch}
 Since we are primarily interested in what happens close to ${\bf k}=0$, we consider only the lower branch polariton and neglect any interband interactions.
We thus introduce field operators $\hat\Psi_{L}({\bf x})$ and $\hat\Psi_{L}^{\dagger}({\bf x})$ and write the Hamiltonian (with  $\hbar = 1$) in the 
mixed representation as  as 
\begin{eqnarray}
\hat {\cal H}_{LP} &=& \int d{ {\bf x}} \hat\Psi_{L}^{\dagger}({\bf x})(\hat\omega_{LP}-i\gamma/2)\hat\psi_L({\bf x})\nonumber\\
 &+& \frac{g}{2}\int d{ {\bf x}}(\hat\Psi_L^{\dagger}({\bf x}))^{2}(\hat\Psi_L({\bf x}))^{2}\nonumber \\ &+&\int d{ {\bf x}} (E({\bf x},t)\hat\Psi_L^{\dagger}({\bf x}) + h.c.)
\end{eqnarray}
where the term $\hat\omega_{LP}(-i\nabla)$ 
describes the lower polariton curve obtained from slab calculations described above.  In doing so, we can incorporate 
site-wise disorder directly into the continuum/field theory equations. 

The field operators obey commutation relations: $[\hat\Psi_{i}({\bf x}),\hat\Psi_{j}^{\dagger}({\bf x}')] = \delta({\bf x}-{\bf x}')\delta_{ij}$.
The physical system itself is in a driven non-equilbrium state and we need to include a decay rate
 $\gamma$ that encompasses  both cavity leakage and non-radiative effects, and an external driving field (i.e. the laser pulse)  $E({\bf x},t)$ as measured
within the cavity. Lastly, $g$ describes the polariton-polariton scattering.  In the calculations that follow, we take these to be phenomenological parameters since 
both $\gamma$ and $E({\bf x},t)$ depend upon the particular experimental situation and microcavity cell. 
 The non-linearity parameter can be approximated by $g =4 \pi\hbar a/m^{*}$ where $a$ is the 
$S$-wave scattering length.   It is difficult to obtain a good estimate for this parameter so we set $g = 0.1$ meV which would correspond to a
polariton/polariton collision frequency of  $g/\hbar = 0.152 {\rm ps}^{-1}$ which seems to a reasonable estimate at low exciton densities.  
We are currently working on a more robust estimate of $g$ based upon exciton/exciton scattering in molecular crystals. \cite{Agranovich:2001ly}
Likewise, we take $\gamma = 0.1 $ meV since this produces a decay time which is short compared to the exciton radiative lifetime but longer than 
the Rabi period of ($\sim 10 $ fs)  that can be estimated from the exciton oscillator strength.  

\begin{figure}[t]
{\includegraphics[width=\columnwidth]{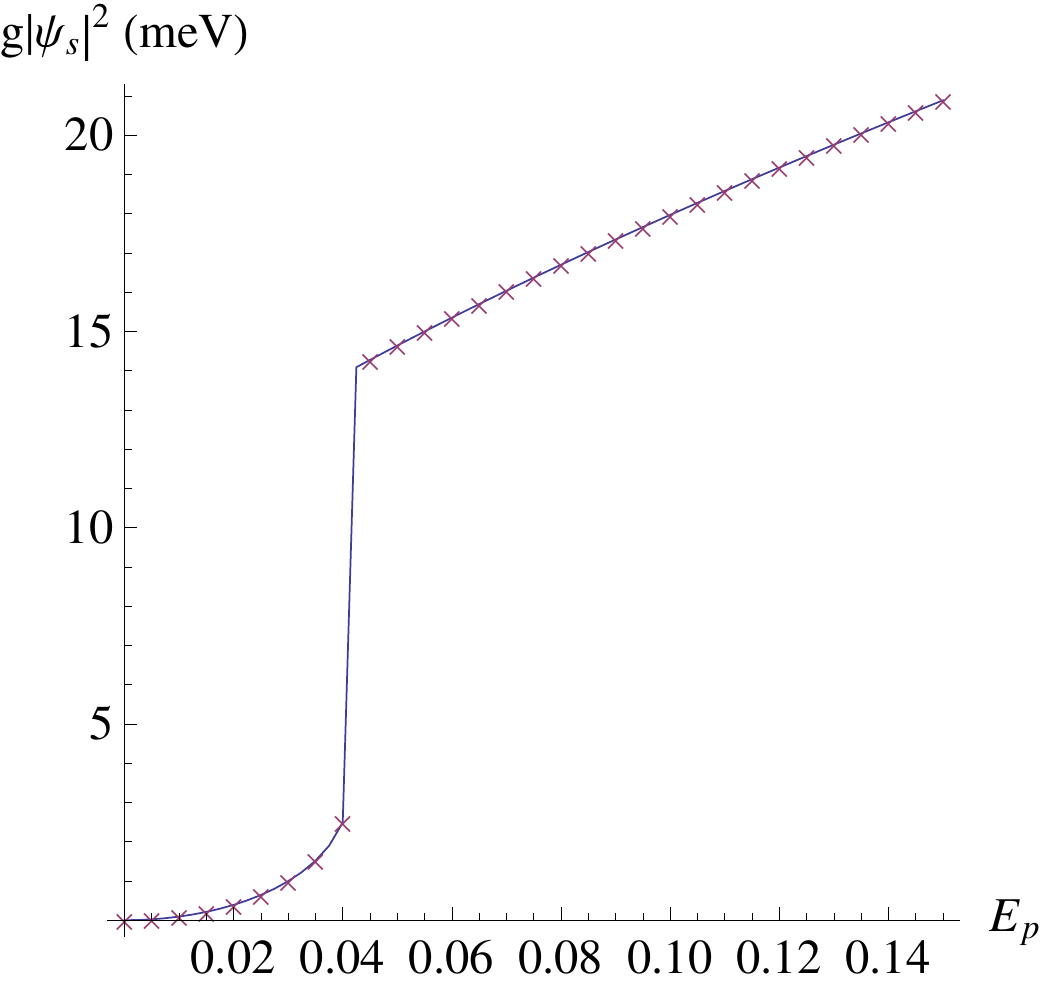}}
\caption{
Mean field energy vs. field strength for $k_{p} = 0.55 \mu {\rm m}^{-1}$.} \label{solns}
\end{figure}

Within the mean-field approximation, the time evolution of the fields $(\psi({\bf x})= \langle \hat \psi_L({\bf x})\rangle)$
is given by\cite{Dickhoff:2008fk,PhysRevLett.93.166401}
\begin{eqnarray}
i \frac{\partial\psi({ {\bf x}})}{\partial t}
 = \left\langle \frac{\delta {\cal H}_{mf}}{\delta \psi_L^{\dagger}({ {\bf x}})} \right\rangle
\end{eqnarray}
This results in the following non-linear, inhomogeneous Schr\"odinger equation:
\begin{eqnarray}
i\frac{ \partial\psi({\bf x})}{\partial t}  &=& 
(\hat\omega_{LP}- i\gamma/2 + g |\psi({\bf x})|^{2})\psi(x) \nonumber \\ &+&E_{pulse}(x;t)
 \end{eqnarray}
 For the case of atomic condensates (and no dissipation or driving fields) this is the Gross-Pitaevskii equation. 
 
We again assume the material to homogeneous and that the laser field can be described as a plane wave within the cavity.
\cite{Carusotto:2006la,PhysRevA.80.043603,PhysRevB.76.115324}
In this case, the steady state equations take the form of a 
Ginzberg-Landau equation of state for the driven system.
\begin{eqnarray}
 (\omega_{LP}({\bf k}_{p})-\omega_{p} - {i}\gamma/2 + g | \psi_{s}|^{2})\psi_{s}  + E_{p} = 0\label{eq-state}
\label{eq:eqofstate}
\end{eqnarray}
In general, the stationary solution $\psi_{s}$ is complex-valued.   In Fig.~\ref{stability}a we plot the mean-field interaction 
energy $ g|\psi_{s}|^{2}$ vs. external field strength $E_{p}$ for $k_{p} = 0.55\mu{\rm m}^{-1}$ which is close to the 
inflection point on the LP dispersion curve.  
Here we see that below a critical value, $E_{p,crit}$, of the field intensity, $ g|\psi_{s}|^{2}$ increases steadily as the field strength increases, jumps sharply before increasing linearly.  

One can consider the pumping intensity to be proportional to the number of polaritons in e Bose gas created by the laser field. 
Consequently, the sudden change in the mean field energy indicates the transition from a ordinary Bose gas to the superfluid or condensate state.  
As we discuss in the Section~\ref{sec:excsp} this transition is an indication that the steady state solution is parametrically unstable over a continuous range of $k$ values either side of $k_{p}$.

\subsection{Excitation spectra }

\label{sec:excsp}
Having determined the steady-state solutions of the homogeneous gas (Eq.~\ref{eq-state}), we need to determine whether or not 
such solutions are stable with regards to imposing a inhomogeneous perturbation.  
For this, we derive the normal modes of the  field to fluctuations about the stationary solution
by  writing the field in terms of stationary term and a fluctuation
$\hat\psi = \psi_{s} + \delta\hat\psi$.\cite{Dickhoff:2008fk,PhysRevLett.93.166401}    Substituting this into the energy functional above and writing 
${\cal H} = {\cal H}_{0} + {\cal H}_{1} + {\cal H}_{2} + \cdots$. 
\begin{widetext}
\begin{eqnarray}
{\cal H}_{0} &=& 
\int d{\bf x} \psi_{s}^{*}({\bf x})(\hat\omega_{LP} - \omega_{p}-i\gamma/2)\psi_{s}({\bf x}) \nonumber \\ &+& \frac{g}{2}\int d{\bf x} |\psi_{s}({\bf x})|^{4} + 
\int d{\bf x} (E({\bf x})\psi_{s}^{*}({\bf x})  + h.c.) \\
{\cal H}_{1}  &=&
\int d{\bf x} \delta\hat\psi^{\dagger}({\bf x})\left[ \left( \hat\omega_{LP}-\omega_{p} - i \gamma/2+ g |\psi_{s}({\bf x})|^{2}\right)\psi_{s}({\bf x})+ E({\bf x})\right]
\delta\hat\psi({\bf x}) \nonumber \\ &+&  h.c. \\
{\cal H}_{2} &=& \int d{\bf x} \delta\hat\psi^{\dagger}({\bf x})(\hat\omega_{LP} - \omega_{p} -i\gamma/2+2g | \psi_{s}({\bf x})|^{2})\delta\hat\psi({\bf x})\nonumber \\
&+&\frac{g}{2}\int d{\bf x}\left( (\psi_{s}^{*}({\bf x}))^{2}(\delta\hat\psi({\bf x}))^{2}  + (\psi({\bf x}))^{2}(\delta\hat\psi^{\dagger}({\bf x}))^{2} \right)
\end{eqnarray}
\end{widetext}
Of these, ${\cal H}_{1} = 0$ for the steady-state solution and ${\cal H}_{0}$ gives the energy of the stationary state.  
Since ${\cal H}_{2}$ is  quadratic in the fluctuation operators, stability of the stationary state can be determined by analyzing the 
normal-mode modes given the eigenvalues of the Boglioubov-de Gennes equation.\cite{Dickhoff:2008fk}
\begin{eqnarray}
(\hat{\cal L}-\omega_{p})
\left[\begin{array}{c} u_{i}({\bf x}) \\ v_{i}({\bf x}) \end{array}\right]  
= 
\epsilon
\left[\begin{array}{c} u_{i}({\bf x}) \\ v_{i}({\bf x}) \end{array}\right],
\end{eqnarray}
where the operator $\hat{\cal L}$ is given by
\begin{widetext}
\begin{eqnarray}
\hat{\cal L } = \left[
\begin{array}{cc}
\hat\omega_{LP}  + g|\psi_{s}|^{2}  - i\gamma/2&  g\psi_{s}^{2}e^{+2i{\bf k}_{p}{\bf x}} \\
-g(\psi_{s}^{*})^{2}e^{-2i{\bf k}_{p}\cdot {\bf x}}           &- \hat\omega_{LP}- g|\psi_{s}|^{2} - i\gamma/2 
\end{array}
\right].
\end{eqnarray}

Since we have assumed the system to be translationally invariant,  wave-vector $k$ remains a ''good'' quantum number and
we can Fourier  transform the eigenvalue problem and write
\begin{eqnarray}
 ({\cal L}_{k}-\omega_{p})\cdot\left[\begin{array}{c}u_{\bf k}^{\pm}\\ v_{\bf k}^{\pm}\end{array}\right] = 
 \omega^{\pm}({\bf k})\left[\begin{array}{c}u_{\bf k}^{\pm}\\ v_{\bf k}^{\pm}\end{array}\right] 
\label{eq:Lkmatrix}
\end{eqnarray}
where
 \begin{eqnarray}
{\cal L}_{k} = \left[
\begin{array}{cc}
\omega_{LP}({\bf k}) + g|\psi_{s}|^{2} - i\gamma/2    &  g\psi_{s}^{2} \\
-g(\psi_{s}^{*})^{2}           & -\omega_{LP}(2{\bf k}_{p}-{\bf k})- g|\psi_{s}|^{2} -i\gamma/2 
\end{array}
\right].
\end{eqnarray}
\end{widetext}

 \begin{figure}[t]
\subfigure[]{\includegraphics[width=0.49\columnwidth]{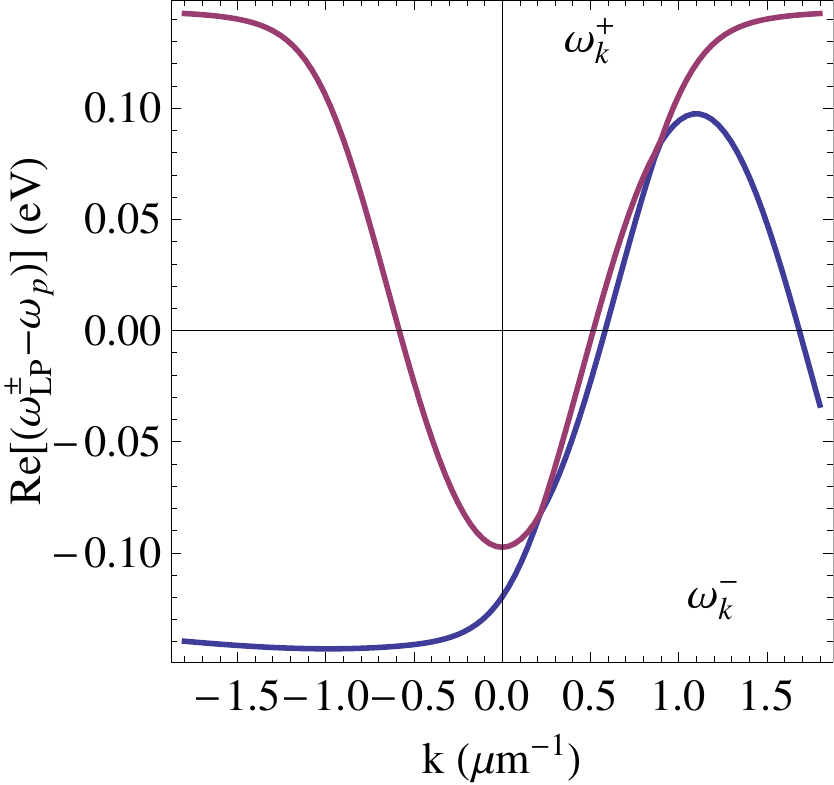}}
\subfigure[]{\includegraphics[width=0.49\columnwidth]{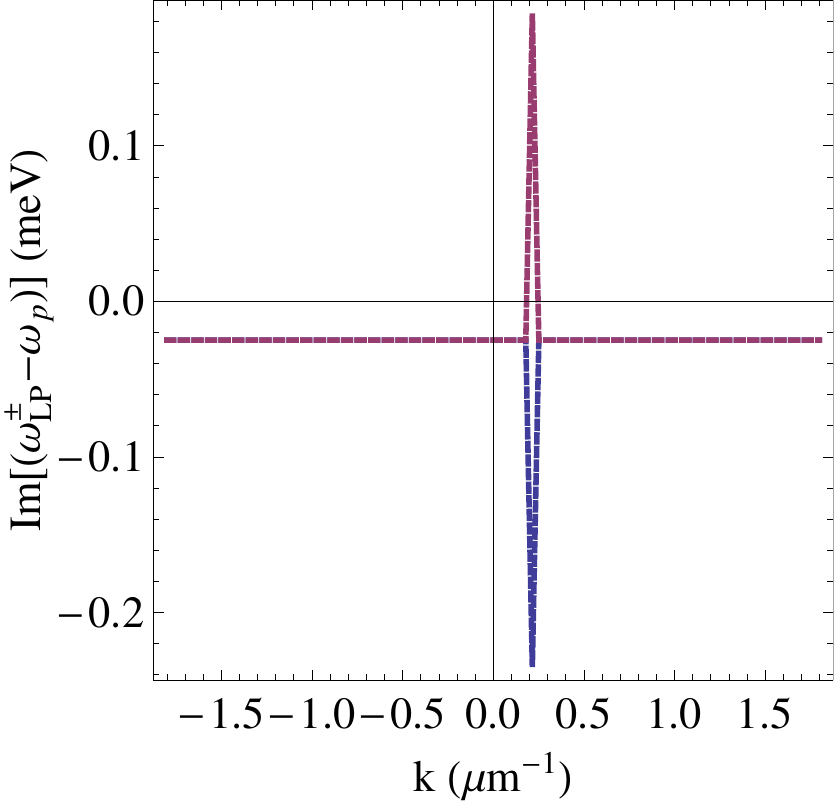}}
\subfigure[]{\includegraphics[width=0.49\columnwidth]{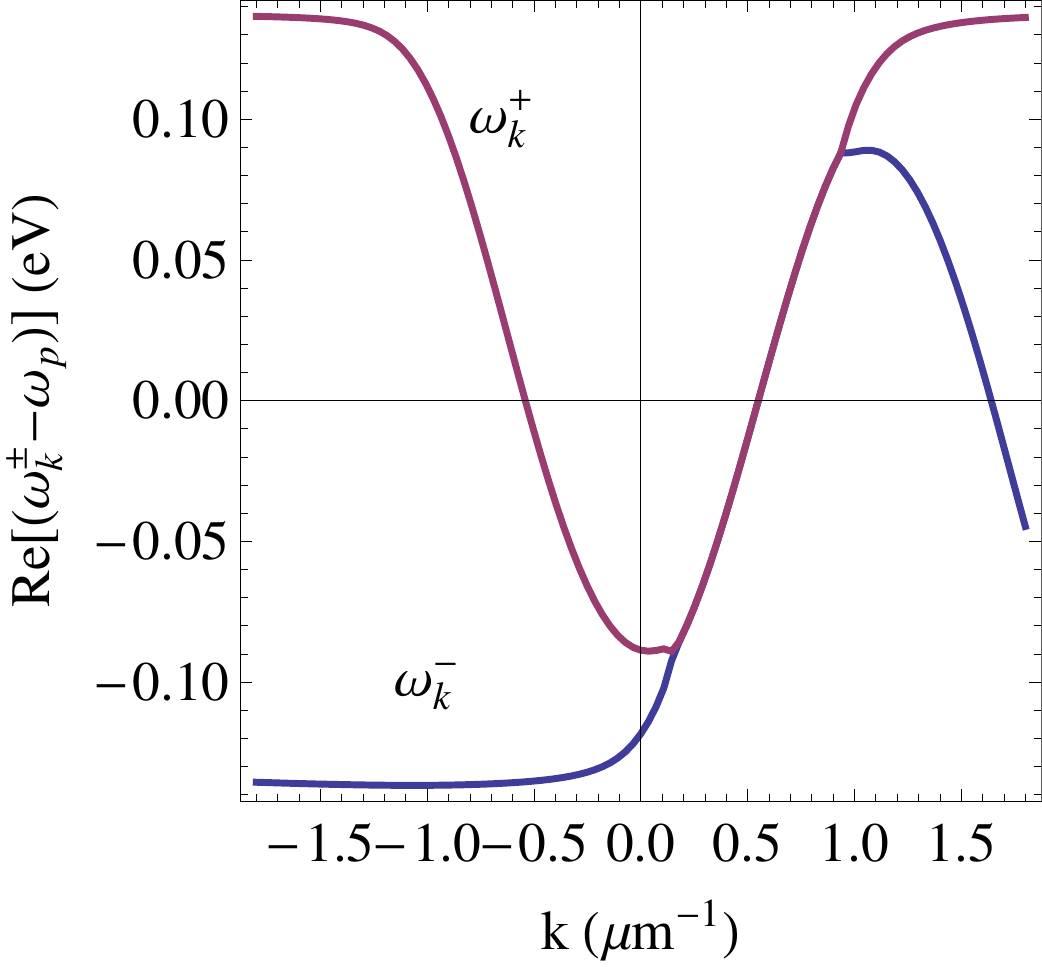}}
\subfigure[]{\includegraphics[width=0.49\columnwidth]{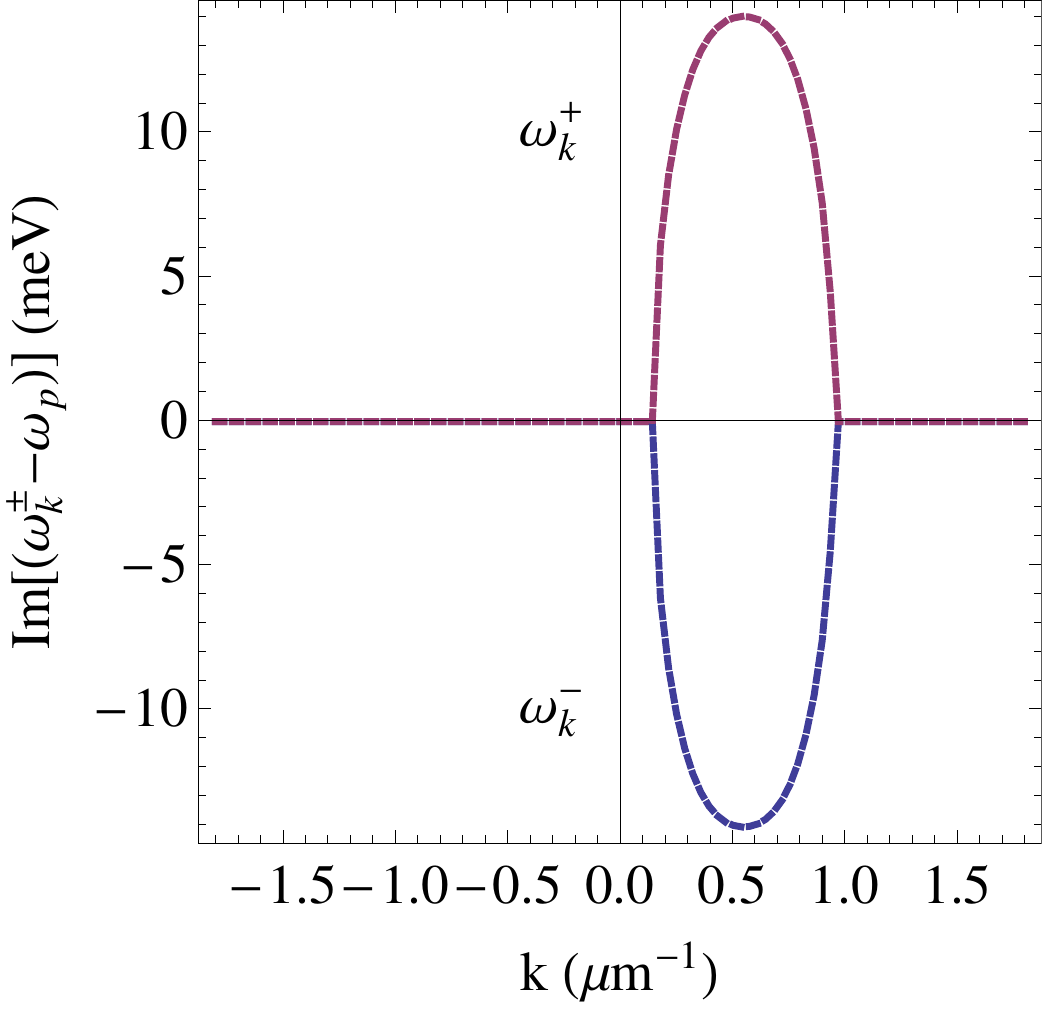}}
\caption{Real and imaginary eigenvalues of ${\cal L}_{k}$ for below (a,b) and above threshold (c,d) $E_{p,crit}$.
 (parameters:   $E_{p}= 0.02 eV$ (a,b),  $E_{p} = 0.05 eV $(c,d), $\omega_{p} = 1.5968eV$, $g = 0.1 meV$, $k_p =0.55 \mu m^{-1}$ along the $x$-direction,and $\gamma = 0.1 meV$) } \label{dispersion2}
\end{figure}

 \begin{figure}[b]
\includegraphics[width=\columnwidth]{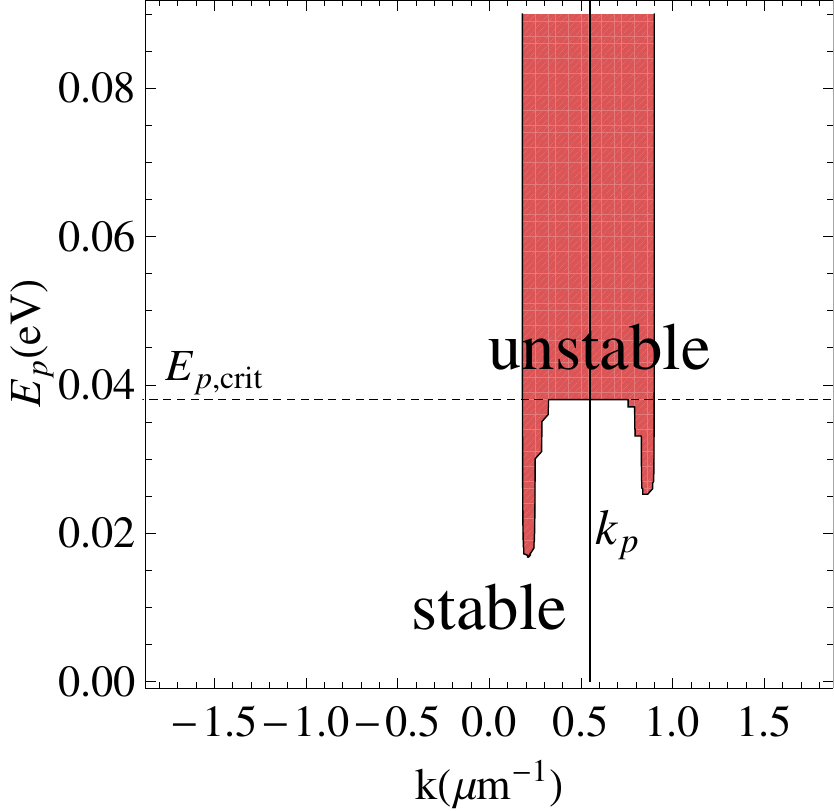}
\caption{
 Mode stability versus applied field strength. 
 Shaded region indicates modes with positive imaginary frequency components. 
Above $E_{p,crit}$  all modes within the branch sticking region are unstable. } \label{stability}
\end{figure}

 \begin{figure}[bb]
\includegraphics[width=\columnwidth]{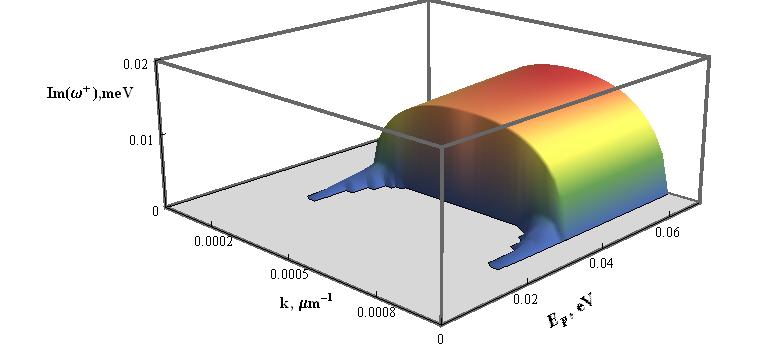}
\caption{Dependence of positive values of $\mathrm{Im}(\omega^+(\mathbf k))$ on the applied filed strength and 
the range of in-plane wave vectors $k$ where system looses its stability.
Flat region represents modes with negative 
imaginary components where steady state is stable (fluctuations' population
decays). However there is a laser regime in which $\mathrm{Im}(\omega^{+})$
becomes positive and population with this mode grows. Here
$\omega_p$=1.5968 eV, $k_p$=0.55 $\mu$m, and $E_{p,crit}$=0.039 eV.}
\label{stability3D}
\end{figure}

At each value of $k$, we find two eigenvalues, $\omega({\bf k})^{\pm}$, which in general are complex valued.  
Those with positive imaginary 
components are unstable modes
meaning that non-equilibrium population within these modes will grow exponentially to form condensate states
while those with negative imaginary components are stable modes and population will decay.\cite{Dickhoff:2008fk} 
The ``idler branch'', $\omega^{-}({\bf k})$, related by symmetry to the ``signal branch'' ($\omega^{+}({\bf k}))$ via 
$\omega^{-}({\bf k}) = 2 \omega_{p} - \omega^{+}(2 {\bf k}_{p}-{\bf k}).$ 
The excitations are are due to the coupling between a generic mode at ${\bf k}$ and an idler mode at $2{\bf k}_{p}-{\bf k}$ induced by 
the driving field. This corresponds to the process where by two photons from the driving field are converted into a signal-idler pair:
$\{{\bf k}_{p},{\bf k}_{p}\} \to \{{\bf k},2{\bf k}_{p}-{\bf k}\}$.
This coupling between modes of different ${\bf k}$ is characteristic of 
 Bogoliubov quasi-particle excitations (bogolons) and parametric amplification.

In Fig.~\ref{dispersion2} we show the eigenvalues of ${\cal L}_{k}$  for the conditions giving rise to Fig.~\ref{solns}. 
Here we see that the idler and signal branches intersect and become degenerate about the points of intersection.
This is known as ``branch sticking''.  Increasing $E_{p}$ causes the branch sticking to spread over the
entire intersection region.  
We also see that  $\Im(\omega^{+}({\bf k})) > 0$ indicating that these modes have become parametrically unstable. 
Increasing $E_{p} > E_{p,crit}$  causes all modes within the intersection region to become unstable.  
In Fig.~\ref{stability}, ~\ref{stability3D} we show the range of $k$ that gives rise to unstable modes.  
For weak field intensity, all modes are stable for all $k$.  As $E_{p}$  increases, 
first one region then a second region becomes unstable.  These eventually 
merge such that above the critical threshold, 
all $\omega^{+}(k)$ within the intersection region have positive imaginary eigenvalues.
It is also important to notice that the energy dispersion is linear (or nearly so) over the entire branch-sticking range.  This 
signals the on-set of a superfluid state according to Landau's criterion.  
As a consequence, non-equilibrium
population within these modes will grow exponentially to form
condensate states.


\begin{figure}[t]
\includegraphics[width=\columnwidth]{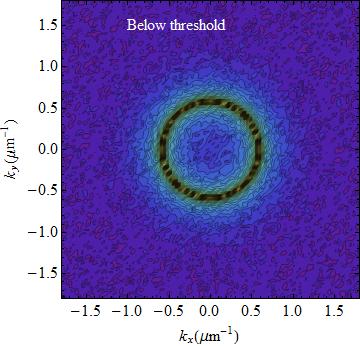}
\caption{Scattered intensity distribution for $k_{p}= 0.55\mu{\rm m}^{-1}$ (plotted as $\log(|\delta\psi({\bf k})|^{2})$) in the Rayleigh scattering regime for $E_p$ below the critical intensity$E_{p,crit}$.}
\label{rayleigh}
\end{figure}

\subsection{Effects of orientational disorder}

We now consider the effects of  orientational
disorder in the film on the critical behavior of the field strength
$E_{p,crit}$ and, consequently, on the stability of the system.
The shape and position of the lower polariton curve, and thus $E_{p,crit}$, 
depends on the  orientation of the molecular units inside the sample.
Even at criogenic temperatures, one expects some degree of disorder in the orientation of the molecular sites
within the lattice.
We can introduce disorder by sampling the orientation of the anthracene molecular unit 
about its crystallographic orientation
(see 
\footnote[33]{The orientation of the molecular $y$-axis of the anthracene molecules 
within the lattice were generated from Gaussian
distribution with the mean fixed at the crystalline value of 
$\bar{\theta}$=20$^\circ$. The standard deviation $\sigma_{\theta}$
about this  from 1$^\circ$ to 15$^\circ$ represents  a measure of the {\em static} disorder within the slab. 
In the simulations presented here, we averaged over 500 realizations of the lattice for a fixed 
$\omega_p$=1.597$eV$ and $k_p$=0.55 $\mu\rm{m}^{-1}$ which corresponds to 
incidence angle of $\approx$ 43$^\circ$ from normal and is close to the inflection point on the LP curve.}
for details).
Since both the exciton dispersion and coupling to the photon field are sensitive to dipoles' orientation,
each sampling will generate a unique LP dispersion. 
We therefore use this to compute $E_{p,crit}$ for different realizations of the lattice for a given angular variance $\sigma_{\theta}$. 

\begin{figure}
\centering
\includegraphics[width=\columnwidth]{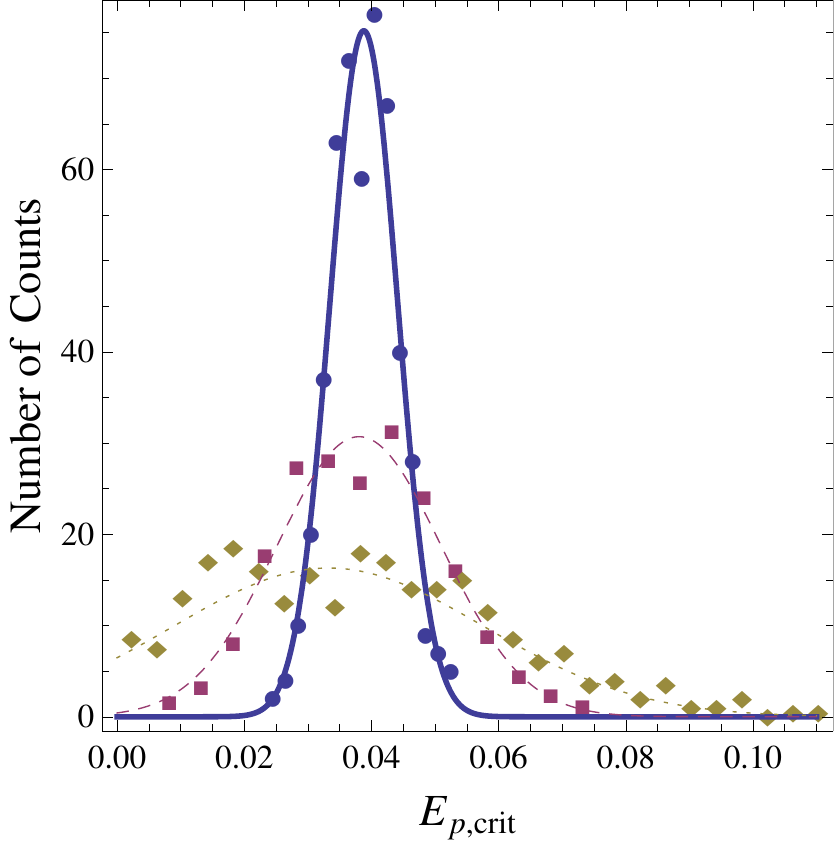}
\caption{\footnotesize{Probability density function of $E_{p,crit}$
    for different degrees of disorder $\delta\theta$;
    $\omega_p$=1.597$eV$, $k_p$ =0.55 $\mu eV$; circles, squares, and
    diamonds correspond to $\delta\theta$=2$^\circ$,
    $\delta\theta$=5$^\circ$ and $\delta\theta$=10$^\circ$
    respectivily. Each set of data is fitted with Gaussian
    distribution shown by solid line. Best fit is obtained by
    Gaussians with: $\mu=0.0388$ and $\sigma=0.0053$ for
    $\sigma_{\theta}=2^\circ$; $\mu=0.0381$ and $\sigma=0.0130$ for
    $\sigma_{\theta}=5^\circ$; $\mu=0.0333$ and $\sigma=0.0245$ for
    $\sigma_{\theta}=10^\circ$.  }}
\label{fig:gaussians}
\end{figure}

\begin{figure}
\centering
\includegraphics[width=\columnwidth]{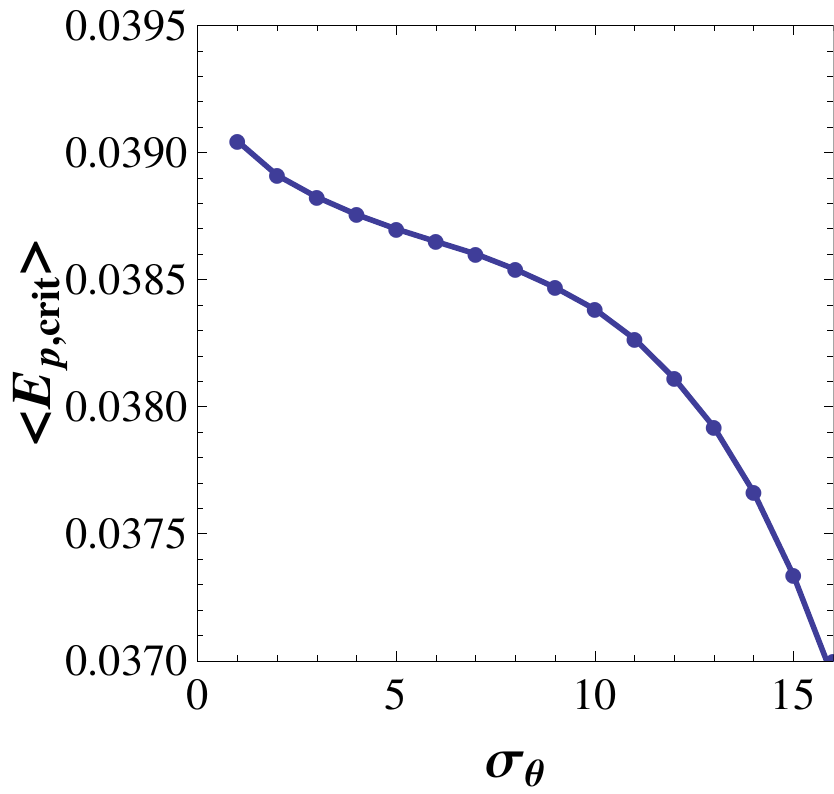}
\caption{\footnotesize{Average of $E_{p,crit}$ over 500 runs versus
    $\delta\theta$. $\omega_p$=1.597$eV$, $k_p$ =0.55 $\mu eV$.}}
\label{fig:average}
\end{figure}

On Fig.~\ref{fig:gaussians} we show the distribution of $E_{p,crit}$  for 3 different values of $\sigma_{\theta}$
(see 
\footnote[34]{The gaussian fits  were determined using built-in \textsf{Mathematica} function \textsf{FindFit[]}.}).
For a perfect crystalline lattice $\sigma_{\theta}=0$ and $E_{p,crit}\approx0.04 eV$.
As one expects, for small $\sigma_{\theta}$  $E_{p,crit}$ is distributed about the $\bar{E}_{p,crit}$ and increasing $\sigma_{\theta}$ results in broader distribution of $E_{p,crit}$.  Fig.~\ref{fig:average} shows values of $\bar{E}_{p,crit}$ for various choices of $\sigma_{\theta}$, averaged over 500 configurations.
Systematic variation of the $E_{p,crit}$ with $\sigma_{\theta}$ is small for low disorder; however, for larger $\sigma_{\theta}$ the behavior of the curve changes and the slope increases.  For a  system held under cryogenic condition one expects $\sigma_{\theta}$ to be very small and consequently we expect that the transition from a normal Bose gas to condensate should be sharp even with some degree of molecular disorder. However, increasing the lattice temperature, the transition will be difficult to observe.



\subsection{Response to inhomogeneous perturbation}

 In the physical system, point defects within the thin film or defects in the reflecting surface introduce potential scattering sites 
 for the polarion field.  In order to examine the effect of such sites we introduce a perturbing field, $f$ and consider its effect on the 
stationary states found above.  Taking $\psi(t)  = \psi_{s} + \delta\psi(t)$ as the time-dependent field in the presence of the scattering sites, this evolves  in time according to
\begin{eqnarray}
i \frac{\partial}{\partial t}\delta\psi = {\cal L}\cdot\delta\psi + f
\end{eqnarray}
As above, we seek stationary solutions to this equation. 
Since the steady-state part satisfies the homogeneous part of this equation, 
 we can consider the (stationary) response to the perturbation by solving the inhomogeneous part
\begin{eqnarray}
\delta\psi({\bf x}) = - {\cal L}^{-1}\cdot f({\bf x})
\end{eqnarray}
Thus, taking $f({\bf x}) = V({\bf x})(\psi_{s},-\psi_{s}^{*})$, we have 
\begin{eqnarray}
\left(
\begin{array}{c}
\delta\psi({\bf k}) \\
\delta \psi^{*}(2{\bf k}_{p}-{\bf k})
\end{array}
\right)
 = -
 {\cal L}_{k}^{-1}\cdot\left(
 \begin{array}{c} \tilde V_{{\bf k}}\psi_{s} \\ - \tilde V_{{\bf k}-2{\bf k}_{p}}\psi_{s}^{*}\end{array}\right)
 \end{eqnarray}
 where $\tilde V_{{\bf k}}$ is the Fourier transform of the perturbing field.  
 In essence, the polaritons are elastically scattered over all  $k$-vectors about the 
 constant energy shell given by $\omega_{LP}({\bf k}_{p})$. 
 The resulting signal is then given by $|\delta \psi({\bf k})|^{2}$.  

In practice, we create a two-dimensional array $(100 \times 100)$  of coordinate grid-points,  select randomly 300 sites
and at those sites set $V({\bf x}) = 0.1 meV$.  At all other sites $V({\bf x})= 0$. The inhomogeneous terms
$\tilde V({\bf k})$ and $\tilde V({\bf k}-2{\bf k}_{p})$ are then found by fast-Fourier transform (FFT). In the latter case, 
we use the shift theorem followed by FFT. In this way, we can rapidly average over multiple realizations of random scattering 
events and compute an averaged signal intensity.

 
We consider the response for a cavity that is off-resonance with the Davydov exciton as depicted in the polariton dispersions in Figs.\ref{dispersion1}.  Here the cavity cutoff is about 0.1eV below the lowest Davydov exciton energy. 
In Fig.~\ref{rayleigh} we show the response signal (plotted as $\log(|\delta\psi({\bf k})|^{2})$)
for the conditions given in Fig.~\ref{dispersion2}(c,d), i.e. close to the LP inflection point. 
The orange/yellow ring indicates that scattering occurs to all $k$-vectors on the energy shell given by 
$\omega_{LP}(k_{s})$.  A polariton droplet created under these conditions would evolve on the ``on-shell'' ring as observed in Ref.\cite{Amo:2009fk}.

\begin{figure}[th]
\subfigure[]{\includegraphics[width=0.49\columnwidth]{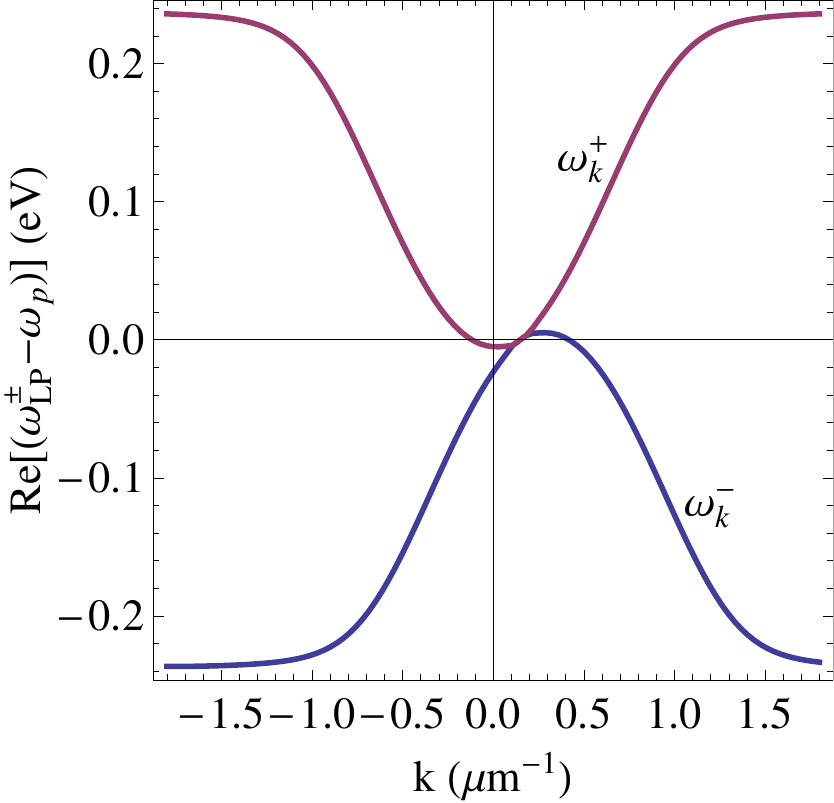}}
\subfigure[]{\includegraphics[width=0.49\columnwidth]{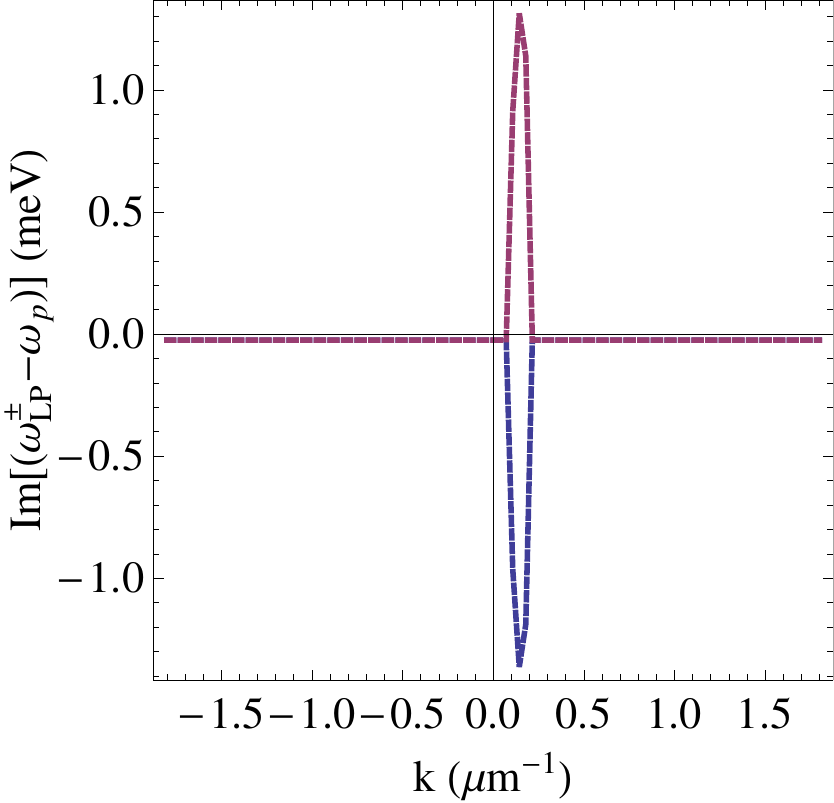}}
\caption{Real and imaginary eigenvalues of ${\cal L}_{k}$.
 (parameters:   $E_{p}= 0.02eV$ (a,b),  $\omega_{p} = \omega_{LP}(k_{p})+1.0 meV$, $g = 0.1 meV$, $k_p =0.15 \mu m^{-1}$,and $\gamma = 0.1 meV$) } \label{dispersion3}
\end{figure}

We next consider what happens when the driving field is close to the bottom of the LP 
dispersion curve and nearly resonant with the LP polariton.  In Fig.~\ref{dispersion3}, the driving 
field is detuned 1meV to the blue from the $\omega_{LP}$.   Here, the idler 
and signal dispersion touch at $k_{p}$ creating a single point of degeneracy that becomes unstable
as $E_{p}$ increases.  

\begin{figure}[t]
\subfigure[]{\includegraphics[width=0.48\columnwidth]{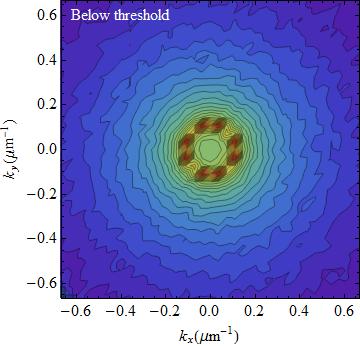}}
\subfigure[]{\includegraphics[width=0.48\columnwidth]{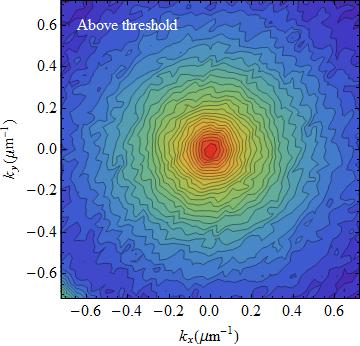}}
\caption{Signal intensity in the super-fluid regime (close to $k=0$) for $E_{p}$ below (a) and above (b) the critical threshold. }\label{BECplot}
\end{figure}

In Fig.~\ref{BECplot}a and b we show the signal resulting from the polaritonic defects.  
 Below threshold, all eigenvalues of ${\cal L}_{k}$ have negative imaginary components and we see a tight ring
indicating that scattering occurs to all $k$-vectors on the $\omega_{LP}(k_{p})$ constant energy shell (Rayleigh scattering).  
Above threshold intensity,  this ring collapses to a single peak at ${\bf k} = \{k_{p},0\}$ indicating the formation 
of a single component ``condensate'' or polariton superfluid state. This does not occur at $k = 0$, rather exactly at
 $k_{p}$.
This is consistent with what one would expect for momentum distribution for an interacting Bose gas \cite{Hugenholtz:1959kx}
$$
n({\bf k}) = z_{o}\delta({\bf k}_{p}) + \tilde n({\bf k})
$$
where $z_{o} = N_{o}/N$ is the condensate fraction and $\tilde n({\bf k})$ is the non-condensate occupation, which is more or less constant with a weak divergence $\sim {\bf k}^{-1}$ as ${\bf k}\to {\bf k}_{p}$. \cite{Gavoret:1964uq}  
This state is special since it is the only one that grows as the number of Bosons in the system $N$  increases.  
Note that the peak is {\em shifted} since the system is driven by the pumping field that has a wave-vector ${\bf k}_{p}$.  
Since we have assumed the pumping field to be of the form $E_{p}e^{i{\bf k}_{p}\cdot {\bf x}}$, 
it equivalent to imposing a moving reference frame on the system.  Furthermore, our theory does not include 
inelastic scattering events that would scattering the signal off the energy-shell. 


\section{Discussion}

In this paper, we discuss the conditions under which polartiton BEC could be achieved in an organic microcavity with $J$-aggregate molecules such as anthracene.
We have identified conditions under which spontaneous coherence and parametric amplification
may be achieved in a polariton gas giving rise to a quantum phase transition to a superfluid phase in an organic 
thin-film system.   This opens the door to experimental probes of polariton condensate dynamics and construction of a quantum phase diagram for organic micro-cavity systems.  To date, evidence for BEC in organic microcavities has not been presented in the literature although efforts to achieve this are underway.

  
It is important to discuss factors not included in our model   which may inhibit or perhaps preclude the formation of stable condensate  states in these systems.  
 Of primary concern is the ability to fabricate well-ordered molecular arrays within the microcavity.  It is 
 absolutely critical that the exciton transition moment be parallel to the reflecting walls of the cavity in order to maximize the coupling to the 
 photon field.  It is also likely that local orientational disorder will inhibit the formation of long-ranged coherences within the 
 cavity. As seen in Fig.~\ref{dispersion1} orientational disorder gives rise to an increase in anti-crossings and an increase in the 
 number of intermediate polariton bands. These may serve to deplete the LP population.   Likewise, strong coupling between the electronic and 
 intramolecular vibrational modes leads to a detuning of the spontaneous emission band (fluorescence) from the vertical absorption absorption band. 
 However, since the strongest electron photon coupling in these systems is local to a given molecular site,
 this will decrease as the Davydov state contributing to the LP is delocalized over multiple molecular sites.   
 Our current approach does not include internal relaxation due to inelastic scattering events.  From detailed balance considerations, 
inelastic scattering will shift the scattering signal towards lower wave vectors and hence the condensate peak to shift towards $k=0$.   
Thermalization must be rapid compared to the polariton lifetime in order to achieve 
 a true BEC state at $k = 0$.  


\begin{acknowledgments}
The work at the University of Houston was funded in part by the National Science Foundation (CHE-1011894) and the Robert A. Welch Foundation (E-1334).
CS acknowledges support from the Canada Research Chair in Organic Semiconductor Materials.
We also wish to acknowledge Dr. Anne Goj (UH) for useful discussions over the course of this work. 
\end{acknowledgments}


\end{document}